\documentclass[a4paper,12pt]{article}

\usepackage[utf8]{inputenc}
\usepackage[T1]{fontenc}
\usepackage{lmodern}

\usepackage{amsmath}
\usepackage[table]{xcolor}
\usepackage{amssymb,amsfonts,amsthm,amscd}
\usepackage{bm}
\usepackage{mathtools}
\usepackage{float}
\usepackage{caption}
\usepackage[subrefformat=parens]{subcaption}

\usepackage{graphicx}
\usepackage{booktabs}
\usepackage{tabularx}
\usepackage{array}

\newcolumntype{Y}{>{\raggedright\arraybackslash}X}

\usepackage{hyperref}

\usepackage{bookmark}            
\hypersetup{
  setpagesize=false,
  bookmarksnumbered=true,
  bookmarksopen=true,
  colorlinks=true,
  linkcolor=blue,
  citecolor=red,
}
\begin{document}

\thispagestyle{empty}
\vspace*{-15mm}

\begin{flushleft}
{\bf OUJ-FTC-26}\\

\end{flushleft}

{\bf }\

\vspace{15mm}

\begin{center}
{\Large\bf
Electroweak Baryogenesis in Top-Philic Type-III Two-Higgs-Doublet Model motivated by the $t\bar{t}$ Excess at the LHC}

\baselineskip 18pt
\vspace{5mm}

Yoshiki Matsuoka

\vspace{4mm}

{\it Nature and Environment, Faculty of Liberal Arts, The Open University of Japan, Chiba 261-8586, Japan\\

}

\end{center}

\vspace{3.5cm}

\begin{flushleft} 
Email: machia1805@gmail.com  
\end{flushleft}

\vspace{7mm}
\begin{center}
\begin{minipage}{14cm}
\baselineskip 16pt
\noindent

\begin{abstract}
The CMS and ATLAS Collaborations have reported a $t\bar{t}$ threshold enhancement possibly involving a pseudoscalar toponium quasi-bound state and an elementary pseudoscalar. Identifying the elementary pseudoscalar with the CP-odd Higgs boson, we investigate electroweak baryogenesis in the top-philic Type-III two-Higgs-doublet model (2HDM), whose scalar sector may support a strong first-order electroweak phase transition. 

We compare two scenarios: explicit CP violation and transitional CP violation (TCPV). In the first, a complex additional top-quark Yukawa coupling induces a spatially varying phase of the top-quark mass across the bubble wall. This generates a chiral asymmetry that electroweak sphalerons convert into baryon asymmetry. In TCPV, CP symmetry would be broken inside the bubble wall but preserved in the vacuum in the absence of explicit CP violation. A small explicit CP-violating bias would favor one of the otherwise CP-conjugate wall configurations, avoiding exact cancellation of the generated baryon asymmetry. This could allow baryogenesis without large explicit CP violation, potentially alleviating electric dipole moment constraints. Combining a one-loop finite-temperature analysis of bubble nucleation with transport calculations, we find that electroweak baryogenesis with explicit CP violation remains possible, although this conclusion depends on the thermal-resummation schemes. No viable TCPV solution is found in the parameter regions explored.

\end{abstract}
\end{minipage}
\end{center}

\vspace{0.5cm}
\newpage
\section{Introduction}
Recently, the CMS and ATLAS Collaborations have reported an excess near the production threshold in the invariant-mass spectrum of top-antitop pairs\cite{topo_1,topo_2,topo_3,topo_4, toponium,topo_5,toponium_atlas}. The possible observation of a pseudoscalar toponium quasi-bound state, which had long been regarded as experimentally elusive, has attracted considerable attention. In addition to an interpretation in terms of a pseudoscalar toponium quasi-bound state, it has also been suggested that an additional elementary pseudoscalar field coexisting with toponium may contribute to the observed excess\cite{topotopo3,topotopo31,topotopo32,topotopo33}.

In this study, we interpret an additional pseudoscalar field as the CP-odd Higgs boson of a two-Higgs-doublet model (2HDM) and investigate electroweak baryogenesis\cite{BAU0,BAU1} within this framework. Depending on the structure of the Yukawa interactions, 2HDMs can be classified into several types. The most commonly studied realizations are the Type-I, Type-II, Type-X, and Type-Y 2HDMs\cite{2hdm}. In these models, discrete symmetries are imposed to specify which Higgs doublet couples to up-type quarks, down-type quarks, and charged leptons, thereby preventing tree-level flavor-changing neutral currents (FCNCs)\cite{2hdm,DM2}.

However, within these conventional Yukawa structures, the coupling of the nonstandard Higgs bosons to top quarks, which plays a central role in interpreting $t\bar{t}$ excess, is subject to stringent experimental constraints\cite{toponium,topotopo31}. We therefore consider a Type-III 2HDM, in which no such discrete symmetry is imposed and both Higgs doublets can, in general, possess independent Yukawa interactions. To suppress potentially dangerous FCNCs, we focus on a top-philic limit in which all nonstandard Yukawa couplings other than the coupling to the top quark are taken to be negligibly small\cite{topphi}.

In studying electroweak baryogenesis in the top-philic Type-III 2HDM, we consider two distinct sources of CP violation (CPV). The first is explicit CP violation, as commonly considered in related previous studies, in which the additional top Yukawa coupling $\rho_{tt}$ is taken to be complex and thus provides an explicit CP-violating source\cite{BAU1,BAU2}.

Motivated by electric dipole moment constraints on explicit CP violation, we also consider transitional CP violation (TCPV)\cite{TCPV1,TCPV2}, in which the dominant CPV source is generated inside the bubble wall. A small imaginary component of a coupling is introduced only to lift the degeneracy between the two CP-conjugate branches and select one of them.

\begin{figure}[t]
  \centering
  \includegraphics[width=0.98\linewidth]{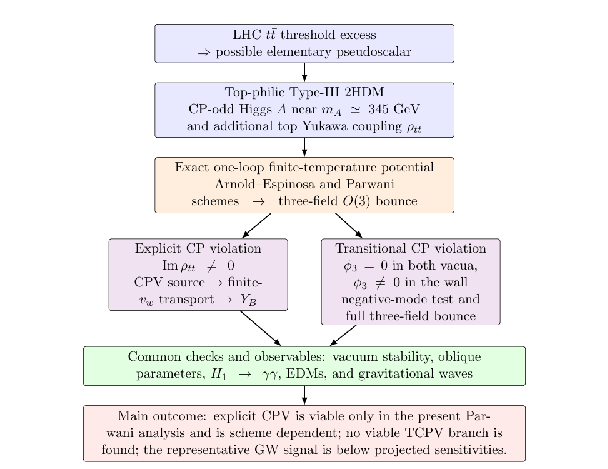}
  \caption{
    Conceptual overview of the motivation, numerical workflow,
    two CP violation scenarios, phenomenological tests, and
    principal conclusions of this work.
  }
  \label{fig:analysis_overview}
\end{figure}

The analysis is designed to answer three questions: whether the top-philic model can support a strong first-order electroweak phase transition near the pseudoscalar mass suggested by the $t\bar t$ threshold excess; whether either CPV mechanism can produce the observed baryon asymmetry after the three-field bounce and transport equations are solved; and whether the surviving points satisfy collider and dipole moment constraints and yield an observable gravitational-wave signal. The logic of the analysis and the principal conclusions are summarized in Fig.~\ref{fig:analysis_overview}.

Within the model setup and parameter regions examined in this work, we find that the explicit CP violation scenario yields candidate parameter points that reproduce the observed baryon asymmetry of the Universe (BAU) when the Parwani prescription\cite{Par} is employed. However, no corresponding viable solution is found with the Arnold--Espinosa prescription\cite{AE}, indicating a significant dependence on the thermal-resummation scheme. In the TCPV scenario, no viable CP-odd bubble-wall branch is identified in the parameter regions explored. We discuss the physical origin of the different outcomes in the two CPV scenarios.

The remainder of this paper is organized as follows: 
In Section \ref{sec2}, we introduce the tree-level model employed in this study and present the one-loop effective potential. In Section \ref{sec3}, we describe the generation of the CP-violating source in the explicit CP violation scenario. We then incorporate this source into the transport equations\cite{Trans} and examine whether the resulting baryon asymmetry $Y_B$ is consistent with the observed value. In Section \ref{sec4}, we explain the mechanism of transitional CP violation\cite{TCPV1,TCPV2} and discuss why it does not lead to successful baryogenesis in the present analysis. In Section \ref{sec5}, we discuss possible connections with gravitational-wave signatures. Finally, our conclusions are summarized in Section \ref{sec6}.

The explicit form of the one-loop effective potential is given in \ref{firstappendix}. The transport equations are summarized in \ref{secondappendix}, while the derivation of the baryon asymmetry $Y_B$ is presented in \ref{thirdappendix}.
In \ref{fourthappendix}, we present the one-loop and leading two-loop contributions to the electric dipole moments of the electron, neutron, and proton.

\section{Top-Philic Type-III Two-Higgs-Doublet Model}
\label{sec2}

In this section, we describe the setup of the top-philic Type III
2HDM considered in this work.

Strictly speaking, the Type-III classification primarily refers to the
structure of the Yukawa interactions.
In the Type-I, Type-II, Type-X, and Type-Y 2HDMs, a discrete
$Z_2$ symmetry is usually imposed to forbid tree-level
flavor-changing neutral currents (FCNCs).
By contrast, no such discrete symmetry is imposed on the Yukawa sector
of the Type-III 2HDM.
Consequently, both Higgs doublets can, in general, couple to the same
fermion species.
The scalar potential may also contain terms that break the $Z_2$
symmetry.

We work in the Higgs basis, in which the two scalar doublets $H_1$ and
$H_2$ satisfy
\begin{align}
    H_1
    &=
    \frac{1}{\sqrt{2}}
    \begin{pmatrix}
        G^{\pm} \\
        \phi_1+v+iG_0
    \end{pmatrix},
    &
    H_2
    &= \begin{pmatrix}
        H^{\pm} \\
        \phi_2+i\phi_3
    \end{pmatrix}.
\end{align}
The most general renormalizable tree-level scalar potential is then
written as
\begin{align}
V_{\mathrm{tree}}
={}&
m_{11}^{2}
\left(H_1^\dagger H_1\right)
+
m_{22}^{2}
\left(H_2^\dagger H_2\right)
+
\left[
m_{12}^{2}
\left(H_1^\dagger H_2\right)
+
\mathrm{h.c.}
\right]
\nonumber\\
&+
\frac{\lambda_1}{2}
\left(H_1^\dagger H_1\right)^2
+
\frac{\lambda_2}{2}
\left(H_2^\dagger H_2\right)^2
+
\lambda_3
\left(H_1^\dagger H_1\right)
\left(H_2^\dagger H_2\right)
\nonumber\\
&+
\lambda_4
\left(H_1^\dagger H_2\right)
\left(H_2^\dagger H_1\right)
\nonumber\\
&+
\left[
\frac{\lambda_5}{2}
\left(H_1^\dagger H_2\right)^2
+
\lambda_6
\left(H_1^\dagger H_1\right)
\left(H_1^\dagger H_2\right)
+
\lambda_7
\left(H_1^\dagger H_2\right)
\left(H_2^\dagger H_2\right)
+
\mathrm{h.c.}
\right].
\label{eq:tree_potential}
\end{align}
In the present analysis, we set\footnote{
A transformation to the Higgs basis allows us to choose
$\langle H_1\rangle \neq 0$ and
$\langle H_2\rangle = 0$, but this basis choice alone does not
imply that both $m_{12}^{2}$ and $\lambda_6$ vanish.
In our sign convention, the tree-level stationary condition
$m_{12}^{2}=-\lambda_6v^2/2$ holds, and hence
$m_{12}^{2}=0$ follows after imposing the alignment condition
$\lambda_6=0$.
For nonzero $\lambda_7$ and $\rho_{tt}$, these terms are generally
regenerated by loop corrections. \cite{BAU2} adopts an effective-potential scheme in which
counterterms cancel the one-loop shifts of the vacuum position and
the curvature matrix evaluated at the vacuum.
For simplicity, we impose the renormalized conditions
$m_{12}^{2}=\lambda_6=0$ at a reference scale and neglect their
radiative regeneration in the present analysis.
}

\begin{align}
    m_{12}^{2} &= 0,
    &
    \lambda_6 &= 0,
\label{eq:model_assumptions}
\end{align}
and take $\lambda_1,\ldots,\lambda_5,\lambda_7$ to be real.

In a general Type-III 2HDM, both Higgs doublets can couple to all
fermion species.
Here, we focus on the top-philic limit, in which only the nonstandard
Yukawa coupling of $H_2$ to the top quark is sizable.
The relevant top-quark Yukawa interactions are therefore given by
\begin{align}
-\mathcal{L}_{Y}
\supset
y_t\,
\overline{Q}_{3L}
\widetilde{H}_1
t_R
+
\rho_{tt}\,
\overline{Q}_{3L}
\widetilde{H}_2
t_R
+
\mathrm{h.c.},
\label{eq:top_yukawa}
\end{align}
where
\begin{align}
    \widetilde{H}_i
    &\equiv
    i\sigma_2 H_i^{*},
    \\
    \rho_{tt}
    &=
    \rho_{ttR}
    +
    i\rho_{ttI}
\label{eq:rhott_definition}
\end{align}

denotes the additional top-quark Yukawa coupling.
To suppress potentially dangerous FCNCs, all non-standard Yukawa
couplings other than $\rho_{tt}$ are assumed to be negligibly small.
\ \\
\ \\
To investigate electroweak baryogenesis in this model, we introduce
the one-loop finite-temperature effective potential.
In the Arnold--Espinosa prescription\cite{AE}, the one-loop finite-temperature
effective potential can be written schematically as
\begin{align}
V_{\mathrm{eff}}^{\mathrm{AE}}
(\boldsymbol{\phi},T)
={}&
V_0(\boldsymbol{\phi})
+
V_{\mathrm{CW}}(\boldsymbol{\phi})
+
V_T(\boldsymbol{\phi},T)
+
V_{\mathrm{ring}}^{\mathrm{AE}}
(\boldsymbol{\phi},T),
\end{align}
where $V_0$, $V_{\mathrm{CW}}$, $V_T$, and
$V_{\mathrm{ring}}^{\mathrm{AE}}$ denote the tree-level potential,
the Coleman--Weinberg potential, the one-loop finite-temperature
correction, and the ring contribution that resums the infrared-sensitive
bosonic zero modes, respectively.

In the Parwani prescription\cite{Par}, by contrast, the thermally resummed masses
are inserted directly into the bosonic Coleman--Weinberg and thermal
contributions. The effective potential is therefore schematically
written as
\begin{align}
V_{\mathrm{eff}}^{\mathrm{P}}
(\boldsymbol{\phi},T)
={}&
V_0(\boldsymbol{\phi})
+
V_{\mathrm{CW}}^{\mathrm{P}}
(\boldsymbol{\phi},T)
+
V_T^{\mathrm{P}}
(\boldsymbol{\phi},T),
\end{align}
without adding a separate ring term.

Instead of employing the commonly used high-temperature expansion, we
numerically evaluate the one-loop thermal functions in their unexpanded
integral forms:
\begin{align}
J_B(y)
&=
\int_0^\infty dq\,
q^2
\ln
\left[
1-\exp\left(-\sqrt{q^2+y}\right)
\right],
\label{eq:thermal_function_boson}
\\
J_F(y)
&=
\int_0^\infty dq\,
q^2
\ln
\left[
1+\exp\left(-\sqrt{q^2+y}\right)
\right],
\label{eq:thermal_function_fermion}
\end{align}
where
\begin{align}
y=\frac{m_i^2(\boldsymbol{\phi})}{T^2}.
\end{align}
To remove the field-independent thermal contributions, we define the
finite-temperature corrections per degree of freedom as
\begin{align}
\mathcal{T}_B(x;T)
&=
\frac{T^4}{2\pi^2}
\left[
J_B\left(\frac{x}{T^2}\right)-J_B(0)
\right],
\\
\mathcal{T}_F(x;T)
&=
-\frac{T^4}{2\pi^2}
\left[
J_F\left(\frac{x}{T^2}\right)-J_F(0)
\right].
\end{align}

We evaluate the thermal functions without using the high-temperature
expansion because, in the field configurations and temperature range
relevant to the phase-transition analysis, the condition
\begin{align}
\frac{\sqrt{\left|m_i^2(\boldsymbol{\phi})\right|}}{T}
\ll 1
\end{align}
is not necessarily satisfied for all field-dependent masses.
Consequently, the accuracy of the high-temperature expansion is not
guaranteed in the parameter region considered here.
We therefore use the unexpanded integral representations given in
Eqs.~\eqref{eq:thermal_function_boson} and
\eqref{eq:thermal_function_fermion}%

Furthermore, to assess the scheme dependence associated with daisy
resummation, we analyze the phase transition using both the
Arnold--Espinosa\cite{AE} and Parwani\cite{Par} prescriptions and examine the robustness
of the resulting phase-transition predictions.

The explicit forms of the field-dependent mass matrices, the thermal
self-energies, and the finite-temperature effective potential in each
resummation prescription are presented in \ref{firstappendix}.

\section{Electroweak Baryogenesis with Explicit CP Violation}\label{sec3}

In this section, we consider a setup with explicit CP violation.
More specifically, we search for the region of coupling parameters
in which, at the nucleation temperature $T_n$, the three-dimensional
Euclidean bounce action $S_3$ satisfies
\begin{align}
    \frac{S_3(T_n)}{T_n}
    \simeq 140,
\end{align}
while the resulting baryon asymmetry is consistent with the observed
value,
\begin{align}
    Y_B^{\mathrm{obs}}
    \simeq
    8.7\times 10^{-11}.
\end{align}

To estimate the theoretical uncertainty associated with the thermal
resummation prescription, we perform the same analysis using both the
Arnold--Espinosa prescription\cite{AE} and the Parwani
prescription\cite{Par}, and examine the robustness of the resulting
predictions.
In the present analysis, we require the above conditions to be
satisfied by identical or nearby combinations of couplings in both
prescriptions.
When a viable solution is found only in one prescription, with no
corresponding viable parameter point in the other, we regard the result
as being strongly dependent on the resummation scheme and exclude it
from the set of robust solutions.

Here, explicit CP violation refers to the case in which the additional
top-quark Yukawa coupling introduced in the previous section,
\begin{align}
    \rho_{tt}
    =
    \rho_{ttR}
    +
    i\rho_{ttI},
\end{align}
has a nonvanishing imaginary part,
\begin{align}
    \rho_{ttI}
    =
    \operatorname{Im}\rho_{tt}
    \neq 0.
\end{align}
This complex phase induces a spatially varying complex phase in the
top-quark mass across the bubble wall and thereby generates the
CP-violating source required for electroweak baryogenesis.

To solve the bounce equations numerically and obtain the corresponding
bubble profiles, we employ \texttt{CosmoTransitions}\cite{tra3}.
For the transport analysis, we use the system of transport equations
presented in \ref{secondappendix}, while the baryon asymmetry
$Y_B$ is evaluated following the procedure described in \ref{thirdappendix}.

For clarity, the common numerical framework and the principal acceptance conditions are collected in Table~\ref{tab:numerical_setup}. 

\begin{table}[t]
\centering
\caption{Summary of the numerical framework and principal acceptance conditions used in the explicit-CPV and TCPV analyses.}
\label{tab:numerical_setup}
\small
\begin{tabularx}{\textwidth}{@{}p{3.2cm}YY@{}}
\toprule
Item & Numerical choice or condition & Role in the analysis \\
\midrule
Scalar-sector assumptions
& Higgs basis; $m_{12}^2=\lambda_6=0$; top-philic limit with only $\rho_{tt}$ sizable
& Defines the model and suppresses nonstandard flavor-changing Yukawa interactions. \\
Finite-temperature potential
& Unexpanded $J_B$ and $J_F$ functions; Arnold--Espinosa and Parwani resummation prescriptions\cite{Par,AE}
& Tests the sensitivity of the phase-transition prediction to thermal resummation. \\
Nucleation and transition strength
& $S_3(T_n)/T_n\simeq140$ and $v_n/T_n\gtrsim1$
& Selects nucleating, strongly first-order transitions. \\
Explicit CPV
& $\operatorname{Im}\rho_{tt}\neq0$; finite-wall-velocity two-moment transport for $t_L,b_L,t_R,h$ at $O(3)$-symmetric three-field bounce in $(\phi_1,\phi_2,\phi_3)$ using \texttt{CosmoTransitions}\cite{tra3}
& Computes the CPV source and the baryon asymmetry $Y_B$. \\
TCPV
& CP-symmetric endpoints; lowest CP-odd eigenvalue $\omega_0^2<0$ followed by a full three-field bounce check
& Tests whether a wall-localized CP-violating branch exists and dominates. \\
Phenomenological constraints
& Vacuum stability, electroweak oblique parameters, $H_1\to\gamma\gamma$ at $2\sigma$, and EDM limits near $m_A\simeq345~\mathrm{GeV}$
& Removes theoretically or experimentally excluded points. \\
Gravitational waves
& $T_*=T_n$; sound-wave plus turbulence contributions; $\Upsilon_{\rm sw}=1$ and $\kappa_{\rm turb}=0.05\kappa_{\rm sw}$
& Provides the representative present-day stochastic GW spectrum. \\
\bottomrule
\end{tabularx}
\end{table}

In the numerical analysis, we first impose the theoretical conditions
for vacuum stability. We further require the electroweak oblique
parameters\cite{2hdm} and the constraint from the
$H_1\to\gamma\gamma$ decay\cite{photon} to be satisfied within their
respective $2\sigma$ ranges. In addition, in the region
$m_A\simeq345~\mathrm{GeV}$, we impose the dipole moment constraints
summarized in \ref{fourthappendix}.

Under these conditions, we find no parameter point in the
Arnold--Espinosa prescription\cite{AE} that simultaneously satisfies
the nucleation condition and the requirement of a strong first-order
electroweak phase transition. By contrast, the Parwani
prescription\cite{Par} yields parameter points for which both
conditions are fulfilled.

In the region where the phase transition is successfully realized in
the Parwani prescription, $\lambda_2$ is not excessively small and can
take values of order unity, up to approximately $1.0$. The coupling
$\lambda_7$ can reach
\begin{align}
    |\lambda_7|
    \lesssim
    0.8.
\end{align}
where $|\lambda_7|$ can't take $0.0$.

Meanwhile, $\lambda_3$ typically takes a relatively large value around
$4.0$, whereas $\lambda_4$ and $\lambda_5$ remain approximately within
\begin{align}
    |\lambda_4|,
    \ |\lambda_5|
    \lesssim
    0.2.
\end{align}
Consequently, the nonstandard scalar spectrum in this region is nearly
degenerate. So, 

\begin{align}
    |\rho_{tt}|
    \lesssim
    0.4.
\end{align}

We additionally impose the strong first-order phase-transition
criterion
\begin{align}
    \frac{v_n}{T_n}
    \gtrsim
    1.
\end{align}
For the parameter points satisfying this condition in the Parwani
prescription\cite{Par}, the nucleation temperature lies in the range
\begin{align}
    113~\mathrm{GeV}
    \lesssim
    T_n
    \lesssim
    117~\mathrm{GeV}.
\end{align}

The successful parameter points, however, require the relatively large
coupling $\lambda_3\simeq4.0$. This may enhance higher-order corrections
in the perturbative expansion and increase the sensitivity to the
thermal resummation prescription. Indeed, in the Arnold--Espinosa
prescription\cite{AE}, we numerically find a critical temperature of
\begin{align}
    T_c
    \simeq
    158~\mathrm{GeV},
\end{align}
but no subsequent solution satisfying both the nucleation and
strong-transition conditions is obtained. This qualitative difference
between the Arnold--Espinosa\cite{AE} and Parwani\cite{Par} prescriptions indicates a
significant dependence of the result on the thermal resummation scheme.

The analysis based on the Parwani prescription\cite{Par} therefore suggests that
the model can realize a strong first-order electroweak phase transition
while satisfying the constraints imposed above. Nevertheless, since no
corresponding solution is found in the Arnold--Espinosa prescription\cite{AE},
the result cannot presently be regarded as scheme independent. A
definitive assessment of the viability of the phase transition requires
the inclusion of higher-order finite-temperature corrections and a more
systematic treatment of thermal resummation.

\section{Electroweak Baryogenesis with Transitional CP Violation}\label{sec4}
Transitional CP violation (TCPV)\cite{TCPV1,TCPV2} is a phenomenon in which the Lagrangian and the vacuum states at both ends of a first-order phase transition are CP conserving, while CP symmetry is spontaneously broken only inside the bubble wall generated during the phase transition. In other words, CP violation does not remain permanently in the present vacuum but appears only in the spatial field profile during the transition.
The defining conditions of wall-localized pure TCPV may be summarized as
\begin{align}
 &\text{a CP-symmetric Lagrangian},\nonumber
 \\
 &\bm{\phi}_{\mathrm{false}}\nonumber
 \ \text{and}\
 \bm{\phi}_{\mathrm{true}}
 \ \text{are CP symmetric},\nonumber
 \\
 &\text{the CP-odd field is nonzero only inside the wall}.\nonumber
\end{align}
In the Higgs basis, the neutral background fields are written as
\begin{align}
 H_1^0(r)&=\frac{\phi_1(r)}{\sqrt{2}},
 \\
 H_2^0(r)&=\frac{\phi_2(r)+\phi_3(r)}{\sqrt{2}}
 =\frac{h_2(r)e^{\theta(r)}}{\sqrt{2}},
\end{align}
where
\begin{align}
 h_2(r)&=\sqrt{\phi_2(r)^2+\phi_3(r)^2},
 \\
 \theta(r)&=\operatorname{atan2}\left(\phi_3(r),\phi_2(r)\right).
\end{align}
In a gauge where $H_1^0$ is real, $\theta$ is the relative phase between the two Higgs doublets.
Under CP,
\begin{align}
 \phi_1(r)&\xrightarrow{CP}\phi_1(r),
 \\
 \phi_2(r)&\xrightarrow{CP}\phi_2(r),
 \\
 \phi_3(r)&\xrightarrow{CP}-\phi_3(r),
 \\
 \theta(r)&\xrightarrow{CP}-\theta(r).
\end{align}
Thus, $\phi_3$, or equivalently $\theta$, parametrizes the CP-odd direction.

For pure TCPV, the endpoints satisfy
\begin{align}
 \phi_3(0)&\simeq\phi_{3,\mathrm{true}}=0,
 \\
 \phi_3(\infty)&=\phi_{3,\mathrm{false}}=0,
\end{align}
whereas inside the wall
\begin{equation}
 \max_r |\phi_3(r)|>0.
\end{equation}
Equivalently,
\begin{equation}
 \theta_{\mathrm{true}},\theta_{\mathrm{false}}\in\{0,\pi\},
 \qquad
 \theta(r)\notin\{0,\pi\}
 \quad\text{for some }r.
\end{equation}

The finite-temperature $O(3)$-symmetric bounce action is
\begin{align}
 S_3(T)
 =4\pi\int_0^\infty d r r^2
 \left[
  \frac12\sum_{i=1}^3
  \left(\frac{d\phi_i}{d r}\right)^2
  +V_{\mathrm{eff}}(\bm{\phi}(r),T)
  -V_{\mathrm{eff}}(\phi_f,T)
 \right].
\label{eq:S3-cartesian-en}
\end{align}
The equations of motion are
\begin{equation}
 \frac{d^2\phi_i}{d r^2}
 +\frac{2}{r}\frac{d\phi_i}{d r}
 =
 \frac{\partial V_{\mathrm{eff}}}{\partial\phi_i},
 \qquad i=1,2,3,
\label{eq:bounce-eom-cartesian-en}
\end{equation}
with boundary conditions
\begin{align}
 \left.\frac{d\phi_i}{d r}\right|_{r=0}&=0,
 \\
 \phi_i(r\to\infty)&=\phi_{i,\mathrm{false}}.
\end{align}
If CP is exact,
\begin{equation}
 \frac{S_3^{(+)}}{T}
 =
 \frac{S_3^{(-)}}{T}.
\label{eq:branch-degeneracy-en}
\end{equation}

Consider a CP-even wall,
\begin{equation}
 \bar{\bm{\phi}}(r)
 =
 \left(\bar\phi_1(r),\bar\phi_2(r),0\right),
\end{equation}
and perturb the CP-odd field as
\begin{equation}
 \phi_3(r)=\eta(r).
\end{equation}
The local curvature in the reduced three-field description is
\begin{equation}
 U_{33}(r;T)
 =
 \left.
 \frac{\partial^2 V_\mathrm{eff}}
 {\partial\phi_3^2}
 \right|_{(\bar\phi_1,\bar\phi_2,0)}.
\label{eq:U33-def-en}
\end{equation}
At our tree-level term,
\begin{equation}
 U_{33}^{(0)}
 =
 m_{22}^2
 +\frac{\lambda_2}{2}\bar\phi_2^2
 +\frac{\lambda_3+\lambda_4-\lambda_5}{2}\bar\phi_1^2
 +\lambda_7\bar\phi_1\bar\phi_2.
\label{eq:U33-tree-en}
\end{equation}
At finite temperature, thermal masses, the Coleman--Weinberg potential, thermal functions, and daisy resummation contribute in addition to \eqref{eq:U33-tree-en}.

A region with \(U_{33}(r;T)<0\) inside the wall indicates that the CP-odd direction tends to become unstable. However, local negativity alone does not establish the existence of a TCPV wall. One must include the gradient energy and study the second variation
\begin{align}
 \delta^2S_3
 =
 2\pi\int_0^\infty d r\,r^2\,
 \eta(r)
 \left[
  -\frac{d^2}{d r^2}
  -\frac{2}{r}\frac{d}{d r}
  +U_{33}(r;T)
 \right]\eta(r).
\label{eq:second-variation-en}
\end{align}
Defining
\begin{equation}
 \mathcal{O}_{\mathrm{odd}}
 =
 -\frac{d^2}{d r^2}
 -\frac{2}{r}\frac{d}{d r}
 +U_{33}(r;T),
\label{eq:odd-operator-en}
\end{equation}
one solves
\begin{equation}
 \mathcal{O}_{\mathrm{odd}}\eta_n
 =
 \omega_n^2\eta_n.
\end{equation}
If the lowest eigenvalue satisfies
\begin{equation}
 \omega_0^2<0,
\label{eq:odd-negative-mode-en}
\end{equation}
the CP-even wall is unstable against a CP-odd deformation, and a three-field branch with nonzero $\phi_3$ may emerge.

A physical TCPV solution must ultimately be established by solving the nonlinear three-field bounce equations and verifying
\begin{align}
 \max_r|\phi_3(r)|&>\phi_{3,\min},
 \\
 \frac{S_3^{(3)}}{T}
 &<
 \frac{S_3^{(2)}}{T}.
\label{eq:3field-dominance-en}
\end{align}
Here, $S_3^{(2)}$ is the action of the CP-even bounce restricted to $\phi_3=0$, while $S_3^{(3)}$ is the action in the full $(\phi_1,\phi_2,\phi_3)$ field space.

In addition, the nucleation rates of the two branches may be written as
\begin{equation}
 \frac{\Gamma_\pm}{V}
 \simeq
 A_\pm(T)
 \exp\left[-\frac{S_3^{(\pm)}(T)}{T}\right].
\label{eq:branch-rates-en}
\end{equation}
Under exact CP symmetry,
\begin{equation}
 S_3^{(+)}=S_3^{(-)},
 \qquad
 A_+=A_-,
 \qquad
 Y_B^{(-)}=-Y_B^{(+)}.
\end{equation}
Therefore, the ensemble-averaged baryon asymmetry vanishes:
\begin{equation}
 \langle Y_B\rangle_{CP}=0.
\end{equation}
A statistical imbalance may occur in a particular realization of the Universe, but a controlled prediction of the sign and magnitude of the BAU generally requires a branch-selection mechanism.

If a small explicit bias generates
\begin{equation}
 \Delta_{S_3}
 \equiv
 \frac{S_3^{(+)}}{T}
 -
 \frac{S_3^{(-)}}{T},
\end{equation}
and the difference between the prefactors is neglected, then
\begin{equation}
 P_+-P_-
 =
 -\tanh\left(\frac{\Delta_{S_3}}{2}\right).
\label{eq:branch-probability-asymmetry-en}
\end{equation}
For $Y_B^{(+)}=Y_0$ and $Y_B^{(-)}=-Y_0$,
\begin{equation}
 Y_B
 =
 -Y_0\tanh\left(\frac{\Delta_{S_3}}{2}\right)
 \simeq
 -\frac{Y_0}{2}\Delta_{S_3},
 \qquad
 |\Delta_{S_3}|\ll1.
\label{eq:net-bau-bias-en}
\end{equation}
If the bias originates from a physical complex coupling, the setup is no longer strictly pure TCPV but is more appropriately described as slightly biased TCPV or wall dynamics with explicit CP violation.

In this work, we
focus on the parameter region in which the mass of the CP-odd scalar
$A$ is approximately
\begin{align}
    m_A \simeq 345~\mathrm{GeV}.
\end{align}
Under this mass requirement, however, we find no parameter point for
which the two-field bounce action $S_3^{(2)}/T$ yields a viable
first-order phase transition while the CP-odd fluctuation satisfies
the negative-mode condition in
Eq.~\eqref{eq:odd-negative-mode-en}.

Satisfying Eq.~\eqref{eq:odd-negative-mode-en} generally requires a
relatively large value of $\lambda_7$, together with a moderately
large value of $\lambda_5$. In the mass region
$m_A\simeq345~\mathrm{GeV}$, however, the scalar-mass relations and
the theoretical and phenomenological constraints imposed in our
analysis significantly restrict the allowed ranges of $\lambda_5$ and $\lambda_7$. Consequently, the conditions required for
transitional CP violation (TCPV) cannot be simultaneously realized
in the mass region considered in this work.
To further investigate the dependence of this conclusion on the
pseudoscalar mass, we also analyzed several mass regions away from the
reference value $m_A\simeq345~\mathrm{GeV}$.
Among these cases, in the lower-mass region
\begin{align}
    200~\mathrm{GeV}
    \lesssim
    m_A
    \lesssim
    300~\mathrm{GeV},
\end{align}
the lowest eigenvalue $\omega_0$ of the CP-odd fluctuation operator
can be brought numerically closer to zero by appropriately varying
$\lambda_4$,$\lambda_5$, and $\lambda_7$.
Nevertheless, we
find
\begin{align}
    \omega_0>0
\end{align}
for all parameter points examined, and hence the CP-odd negative mode
required for TCPV does not develop. These results indicate that the
realization of TCPV is difficult in this model.

\section{Gravitational-Wave Spectrum}\label{sec5}

A first-order electroweak phase transition can generate a stochastic
gravitational-wave background through the bulk motion of the plasma\cite{gw1}.
The present-day gravitational-wave energy-density spectrum is defined by
\begin{align}
    \Omega_{\rm GW,0}(f)
    \equiv
    \frac{1}{\rho_c}
    \frac{\mathrm{d}\rho_{\rm GW,0}}{\mathrm{d}\ln f},
    \label{eq:omega_gw_definition}
\end{align}
where
\begin{align}
    \rho_c
    =
    \frac{3H_0^2}{8\pi G}
\end{align}
is the present critical energy density.
We present our results in terms of $h^2\Omega_{\rm GW,0}(f)$, where
\begin{align}
    H_0
    =
    100h~{\rm km\,s^{-1}\,Mpc^{-1}}.
\end{align}

In the present analysis, we include the contributions from sound waves
and magnetohydrodynamic turbulence:
\begin{align}
    h^2\Omega_{\rm GW,0}(f)
    =
    h^2\Omega_{\rm sw,0}(f)
    +
    h^2\Omega_{\rm turb,0}(f).
    \label{eq:total_gw_spectrum}
\end{align}
We neglect the contribution from bubble-wall collisions, as appropriate
for the non-runaway phase transitions considered in this work\cite{gw1}.

The gravitational-wave spectrum is characterized by the transition
temperature $T_*$, the transition strength $\alpha$, the inverse
duration $\beta/H_*$, the bubble-wall velocity $v_w$, and the effective
number of relativistic degrees of freedom $g_*$. In the present
calculation, we identify the characteristic transition temperature with
the nucleation temperature:
\begin{align}
    T_*
    =
    T_n.
    \label{eq:gw_transition_temperature}
\end{align}

To define the transition strength, we introduce the free-energy
difference between the false and true vacua,
\begin{align}
    \Delta V(T)
    \equiv
    V_{\rm false}(T)
    -
    V_{\rm true}(T),
\end{align}
and the radiation energy density,
\begin{align}
    \rho_{\rm rad}(T)
    =
    \frac{\pi^2}{30}g_*T^4.
\end{align}
We use the energy-density definition of the strength parameter,
\begin{align}
    \alpha
    \equiv
    \frac{
        \displaystyle
        \Delta V(T_*)
        -
        T_*
        \left.
        \frac{\mathrm{d}\Delta V(T)}{\mathrm{d}T}
        \right|_{T=T_*}
    }{
        \rho_{\rm rad}(T_*)
    }.
    \label{eq:gw_alpha_definition}
\end{align}
Thus, $\alpha$ measures the released transition energy relative to the
radiation energy density of the plasma. A larger value of $\alpha$
generally corresponds to a larger fraction of energy being available
for bulk fluid motion and gravitational-wave production.

The inverse duration of the transition is obtained from the temperature
dependence of the three-dimensional Euclidean bounce action $S_3(T)$.
Since the nucleation rate behaves approximately as
\begin{align}
    \Gamma(T)
    \propto
    \exp\left[
        -\frac{S_3(T)}{T}
    \right],
\end{align}
the inverse-duration parameter in a radiation-dominated Universe is
\begin{align}
    \frac{\beta}{H_*}
    =
    T_*
    \left.
    \frac{\mathrm{d}}{\mathrm{d}T}
    \left(
        \frac{S_3(T)}{T}
    \right)
    \right|_{T=T_*}.
    \label{eq:gw_beta_definition}
\end{align}
A larger value of $\beta/H_*$ corresponds to a shorter phase transition
and a smaller characteristic bubble size\cite{gw2}. Consequently, increasing
$\beta/H_*$ shifts the gravitational-wave spectrum toward higher
frequencies while suppressing its amplitude through the factor
\begin{align}
    \frac{H_*}{\beta}
    =
    \left(
        \frac{\beta}{H_*}
    \right)^{-1}.
\end{align}

The present-day peak frequency of the sound-wave contribution\cite{gw1} is
\begin{align}
    f_{\rm sw}
    =
    1.9\times10^{-5}~{\rm Hz}\,
    \frac{1}{v_w}
    \left(
        \frac{\beta}{H_*}
    \right)
    \left(
        \frac{T_*}{100~{\rm GeV}}
    \right)
    \left(
        \frac{g_*}{100}
    \right)^{1/6}.
    \label{eq:sound_wave_peak_frequency}
\end{align}
Defining
\begin{align}
    x_{\rm sw}
    \equiv
    \frac{f}{f_{\rm sw}},
\end{align}
the sound-wave spectral shape is
\begin{align}
    S_{\rm sw}(f)
    =
    x_{\rm sw}^3
    \left(
        \frac{7}{
            4+3x_{\rm sw}^2
        }
    \right)^{7/2}.
    \label{eq:sound_wave_spectral_shape}
\end{align}
The corresponding present-day spectrum is
\begin{align}
    h^2\Omega_{\rm sw,0}(f)
    ={}&
    2.65\times10^{-6}\,
    \Upsilon_{\rm sw}
    \left(
        \frac{H_*}{\beta}
    \right)
    \left(
        \frac{
            \kappa_{\rm sw}\alpha
        }{
            1+\alpha
        }
    \right)^2
    \nonumber\\
    &\times
    \left(
        \frac{100}{g_*}
    \right)^{1/3}
    v_w
    S_{\rm sw}(f).
    \label{eq:sound_wave_gw_spectrum}
\end{align}
Here, $\kappa_{\rm sw}$ denotes the fraction of the released transition
energy transferred to bulk fluid motion\cite{gw3}. For the numerical scan, we use
the simple non-runaway estimate
\begin{align}
    \kappa_{\rm sw}
    =
    v_w
    \frac{
        \alpha
    }{
        0.73
        +
        0.083\sqrt{\alpha}
        +
        \alpha
    }.
    \label{eq:sound_wave_efficiency}
\end{align}
The factor $\Upsilon_{\rm sw}$ accounts for the finite lifetime of the
acoustic source. In the present scan, we set
\begin{align}
    \Upsilon_{\rm sw}
    =
    1,
\end{align}
corresponding to the long-lived sound-wave approximation.

The present-day characteristic frequency of the turbulence contribution
is
\begin{align}
    f_{\rm turb}
    =
    2.7\times10^{-5}~{\rm Hz}\,
    \frac{1}{v_w}
    \left(
        \frac{\beta}{H_*}
    \right)
    \left(
        \frac{T_*}{100~{\rm GeV}}
    \right)
    \left(
        \frac{g_*}{100}
    \right)^{1/6}.
    \label{eq:turbulence_peak_frequency}
\end{align}
The redshifted Hubble frequency is
\begin{align}
    h_*
    =
    1.65\times10^{-5}~{\rm Hz}\,
    \left(
        \frac{T_*}{100~{\rm GeV}}
    \right)
    \left(
        \frac{g_*}{100}
    \right)^{1/6}.
    \label{eq:redshifted_hubble_frequency}
\end{align}
Defining
\begin{align}
    x_{\rm turb}
    \equiv
    \frac{f}{f_{\rm turb}},
\end{align}
the turbulence spectral shape is
\begin{align}
    S_{\rm turb}(f)
    =
    \frac{
        x_{\rm turb}^3
    }{
        \left(
            1+x_{\rm turb}
        \right)^{11/3}
        \left(
            1+\dfrac{8\pi f}{h_*}
        \right)
    }.
    \label{eq:turbulence_spectral_shape}
\end{align}
The present-day turbulence spectrum is then
\begin{align}
    h^2\Omega_{\rm turb,0}(f)
    ={}&
    3.35\times10^{-4}
    \left(
        \frac{H_*}{\beta}
    \right)
    \left(
        \frac{
            \kappa_{\rm turb}\alpha
        }{
            1+\alpha
        }
    \right)^{3/2}
    \nonumber\\
    &\times
    \left(
        \frac{100}{g_*}
    \right)^{1/3}
    v_w
    S_{\rm turb}(f),
    \label{eq:turbulence_gw_spectrum}
\end{align}
where we adopt
\begin{align}
    \kappa_{\rm turb}
    =
    0.05\,\kappa_{\rm sw}.
    \label{eq:turbulence_efficiency}
\end{align}

The characteristic frequencies scale approximately as
\begin{align}
    f_{\rm peak}
    \propto
    \frac{1}{v_w}
    \left(
        \frac{\beta}{H_*}
    \right)
    T_*,
\end{align}
whereas the amplitudes are proportional to $H_*/\beta$ and increase
with the transition strength through
$\kappa\alpha/(1+\alpha)$. Consequently, a shorter transition with a
larger $\beta/H_*$ produces a higher-frequency but typically smaller
gravitational-wave signal, while a stronger transition with a larger
$\alpha$ generally enhances the signal amplitude.

\begin{figure}[t]
    \centering
    \includegraphics[width=0.86\linewidth]{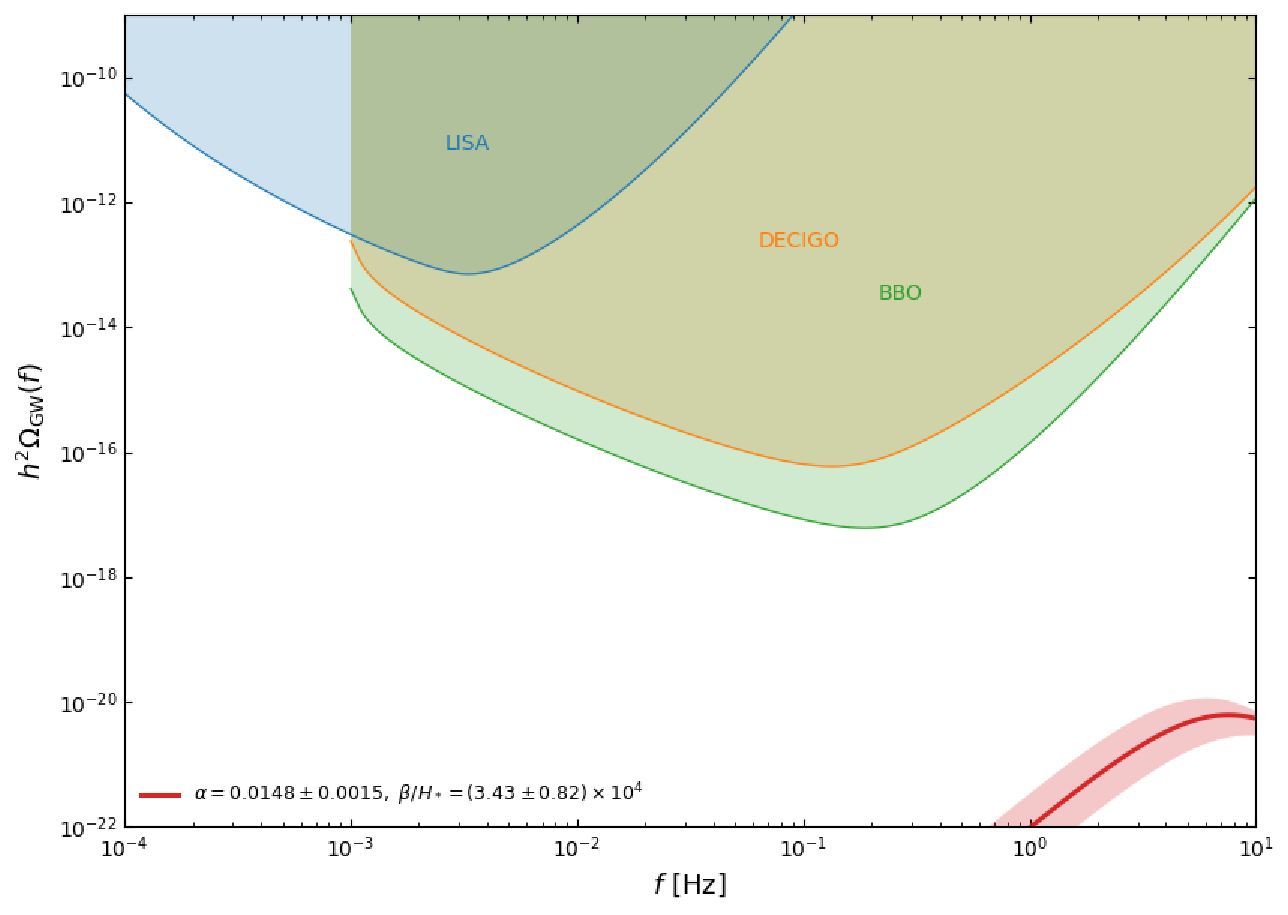}
    \caption{
    Present-day gravitational-wave spectrum for a representative
    parameter point obtained in the Parwani prescription in
    Section~\ref{sec3}.
    The red solid curve represents the sum of the sound-wave and
    turbulence contributions, while the LISA, DECIGO, and BBO curves
    show their projected sensitivities\cite{gw4,gw5}.
    }
    \label{fig:gw_parwani}
\end{figure}

For the representative parameter point obtained in the Parwani
prescription\cite{Par} in Section~\ref{sec3}, the present-day
gravitational-wave spectrum is shown in
Fig.~\ref{fig:gw_parwani}.
The red solid curve represents the sum of the sound-wave and turbulence
contributions, while the LISA, DECIGO, and BBO curves indicate their projected
sensitivities\cite{gw4,gw5}.
For the parameter point shown in the figure, the phase-transition
parameters are
\begin{align}
    \alpha
    =
    0.0148\pm 0.0015,
    \qquad
    \frac{\beta}{H_*}
    =
    (3.43\pm0.82)\times10^{4}.
\end{align}
where transport equations' $v_\omega$ takes $0.1$.
The relatively large value of $\beta/H_*$ corresponds to a
short-duration phase transition. It shifts the peak of the
gravitational-wave spectrum toward higher frequencies while strongly
suppressing its amplitude through the factor $H_*/\beta$.

As shown in Fig.~\ref{fig:gw_parwani}, the predicted spectrum for this
parameter point lies several orders of magnitude below the projected
sensitivities of LISA, DECIGO, and BBO displayed in the figure.
The signal is therefore not expected to be observable with these
baseline detector configurations.

Nevertheless, a substantial future improvement in sensitivity,
particularly in the decihertz-to-few-hertz frequency range, may allow
related regions of the phase-transition parameter space to be probed.
The observation of such a stochastic gravitational-wave background
would not by itself establish electroweak baryogenesis, but it could
provide independent and complementary evidence for the strong
first-order electroweak phase transition required in the present
baryogenesis scenario.

\section{Summary and Discussion}\label{sec6}
In this work, motivated by the excess reported near the
$t\bar{t}$ production threshold at the LHC, we investigated
electroweak baryogenesis in a top-philic Type-III 2HDM.
In particular, we considered two scenarios involving distinct
origins of CP violation and compared their finite-temperature phase
transitions, CPV sources, transport dynamics, and resulting
baryon asymmetries.

In the first scenario, explicit CP violation is introduced through a
nonvanishing imaginary part of the additional top-quark Yukawa coupling.
Within the parameter region considered in this work, the Parwani
prescription\cite{Par} yields parameter points that simultaneously
realize a strong first-order electroweak phase transition and generate
a baryon asymmetry consistent with the observed value.
The complex coupling $\rho_{tt}$ can therefore provide the
CPV source required for electroweak baryogenesis in the
top-philic Type-III 2HDM.

For the same parameter region, however, we find no corresponding
solution in the Arnold--Espinosa prescription\cite{AE} that
simultaneously satisfies the nucleation condition and the requirement
of a strong first-order phase transition.
This result shows that the phase-transition prediction in the
explicit CP violation scenario is currently strongly dependent on the
thermal resummation scheme.
Consequently, the results obtained using the Parwani prescription\cite{Par}
alone do not establish the realization of electroweak baryogenesis in
this model in a scheme-independent manner.

As a second scenario, we considered transitional CP violation
(TCPV), in which a CP-odd field configuration is dynamically generated
inside the bubble wall by finite-temperature effects\cite{TCPV1,TCPV2}.
In the region motivated by the $t\bar{t}$ threshold excess, we found no parameter point that develops the CP-odd negative mode
required for TCPV while maintaining a viable first-order phase
transition.
We also examined several pseudoscalar-mass regions away from
$m_A\simeq345~\mathrm{GeV}$.
Although the lowest eigenvalue of the CP-odd fluctuation operator can
be brought numerically closer to zero in some of these regions, it
remains positive throughout the parameter space in this work.
Within the one-loop finite-temperature treatment and the parameter
regions examined here, the realization of TCPV in the top-philic
Type-III 2HDM is therefore found to be difficult. Therefore, within the model setup and parameter region considered in this work, 
the realization of electroweak baryogenesis appears to require an explicit source of CP violation.

For a representative parameter point obtained in the explicit CPV
scenario using the Parwani prescription\cite{Par}, we also evaluated the gravitational-wave spectrum associated with the
first-order phase transition.

The analyses performed in this work also reveal that sizable
theoretical uncertainties remain in the finite-temperature prediction
of the model.
In particular, the parameter points yielding a successful phase
transition in the Parwani prescription typically require a relatively
large quartic coupling, $\lambda_3\simeq4.0$.
Such a large coupling may enhance higher-order corrections in the
perturbative expansion and may be responsible, at least in part, for
the substantial difference between the Arnold--Espinosa and Parwani
prescriptions\cite{Par,AE}.
A definitive assessment of the explicit CPV scenario therefore
requires an analysis including higher-order finite-temperature
corrections and a more systematic treatment of thermal resummation.

In the TCPV scenario, higher-order corrections may also modify the
CP-odd fluctuation operator and the corresponding bubble-wall solution.
A more reliable determination of whether CP violation can be
dynamically generated inside the wall therefore requires an analysis
beyond the present one-loop treatment.

A relatively large quartic coupling such as $\lambda_3$ can also
accelerate the renormalization-group evolution and may generate a
Landau pole at a scale significantly below the Planck scale.
A dedicated renormalization-group analysis is required to determine
this scale quantitatively.
If perturbativity is lost at a comparatively low energy scale, the
top-philic Type-III 2HDM considered here should be interpreted as an
effective field theory valid only below that scale.
In such a case, a new theoretical structure providing a UV completion
of the model would be required at higher energies.

In summary, potentially nontrivial connections may exist among the
LHC $t\bar{t}$ threshold excess, the electroweak phase transition,
the baryon asymmetry of the Universe, electric dipole moments, and
stochastic gravitational waves.
It demonstrates that possible new physics near the $t\bar{t}$ threshold may raise questions extending
well beyond collider phenomenology, including the dynamics of the
early-Universe phase transition, the origin of the baryon asymmetry,
and the ultraviolet completion of the theory.

\section*{Acknowledgments}
We thank  Shiro Komata, So Katagiri, and Akio Sugamoto for many helpful comments and advices. 

\appendix
\renewcommand{\thesection}{Appendix \Alph{section}}
\section{Each Mass Matrix and One-loop Effective Potentials}\label{firstappendix}
The field-dependent neutral scalar mass is
\begin{equation}
\mathcal{M}_N^2(\bm{\phi})=
\begin{pmatrix}
M_{11} & M_{12} & M_{13} & M_{14}\\
M_{12} & M_{22} & M_{23} & M_{24}\\
M_{13} & M_{23} & M_{33} & M_{34}\\
M_{14} & M_{24} & M_{34} & M_{44}
\end{pmatrix},
\label{eq:neutral_mass_matrix}
\end{equation}
where
\begin{align}
M_{11}
={}&m_{11}^2
+\frac{3}{2}\lambda_1\phi_1^2
+\frac{\lambda_3+\lambda_4}{2}(\phi_2^2+\phi_3^2)
+\frac{\lambda_5}{2}(\phi_2^2-\phi_3^2),
\\[1mm]
M_{12}
={}&(\lambda_3+\lambda_4+\lambda_5)\phi_1\phi_2
+\frac{3}{2}\lambda_{7}\phi_2^2
+\frac{1}{2}\lambda_{7}\phi_3^2,
\\
M_{13}
={}&\lambda_5\phi_2\phi_3,
\\
M_{14}
={}&(\lambda_3+\lambda_4-\lambda_5)\phi_1\phi_3
+\lambda_{7R}\phi_2\phi_3,
\\[1mm]
M_{22}
={}&m_{22}^2
+\frac{3}{2}\lambda_2\phi_2^2
+\frac{1}{2}\lambda_2\phi_3^2
\nonumber\\
&+\frac{\lambda_3+\lambda_4+\lambda_5}{2}\phi_1^2
+3\lambda_{7}\phi_1\phi_2,
\\
M_{23}
={}&\lambda_5\phi_1\phi_3
+\lambda_{7}\phi_2\phi_3,
\\
M_{24}
={}&\lambda_2\phi_2\phi_3
+\lambda_{7}\phi_1\phi_3,
\\[1mm]
M_{33}
={}&m_{11}^2
+\frac{1}{2}\lambda_1\phi_1^2
+\frac{\lambda_3+\lambda_4}{2}(\phi_2^2+\phi_3^2)
-\frac{\lambda_5}{2}(\phi_2^2-\phi_3^2),
\\
M_{34}
={}&\lambda_5\phi_1\phi_2
+\frac{1}{2}\lambda_{7}\phi_2^2
+\frac{3}{2}\lambda_{7}\phi_3^2,
\\
M_{44}
={}&m_{22}^2
+\frac{1}{2}\lambda_2\phi_2^2
+\frac{3}{2}\lambda_2\phi_3^2
\nonumber\\
&+\frac{\lambda_3+\lambda_4-\lambda_5}{2}\phi_1^2
+\lambda_{7}\phi_1\phi_2.
\label{eq:neutral_mass_entries}
\end{align}
\ \\
\ \\

The field-dependent charged scalar
mass matrix is
\begin{equation}
\mathcal{M}_C^2(\bm{\phi})=
\begin{pmatrix}
A & C\\
C^* & D
\end{pmatrix},
\label{eq:charged_mass_matrix}
\end{equation}
with
\begin{align}
A
={}&m_{11}^2
+\frac{\lambda_1}{2}\phi_1^2
+\frac{\lambda_3}{2}(\phi_2^2+\phi_3^2),
\\
D
={}&m_{22}^2
+\frac{\lambda_2}{2}(\phi_2^2+\phi_3^2)
+\frac{\lambda_3}{2}\phi_1^2
+\lambda_{7}\phi_1\phi_2,
\\
C
={}&\frac{1}{2}\left[
\lambda_5\phi_1(\phi_2+i\phi_3)
+\lambda_4\phi_1(\phi_2-i\phi_3)
+\lambda_7(\phi_2^2+\phi_3^2)
\right].
\end{align}
So, each thermal resummation matrix is
\begin{equation}
\Pi=
\begin{pmatrix}
\Pi_{11}& \Pi_{12}&\Pi_{13}&\Pi_{14}\\
\Pi_{21}&\Pi_{22}&\Pi_{23}&\Pi_{24}\\
\Pi_{31}&\Pi_{32}&\Pi_{11}&\Pi_{34}\\
\Pi_{41}&\Pi_{42}&\Pi_{43}&\Pi_{22}
\end{pmatrix}
\end{equation}
and 
\begin{equation}
\Pi^{charged}=
\begin{pmatrix}
\Pi^{charged}_{11}& \Pi^{charged}_{12}\\
\Pi^{\ast charged}_{21}&\Pi^{charged}_{22}
\end{pmatrix}
\end{equation}
where
\begin{align}
\Pi_{11}={}&\Pi_{33}=\frac{\lambda_1}4T^2+\frac{\lambda_3}6T^2+\frac{\lambda_4}{12}T^2+\frac{g^2_2}8T^2+\frac{(g_Y^2+g_2^2)}{16}T^2+\frac{y^2_t}4T^2,
\\
\Pi_{22}={}&\Pi_{44}=\frac{\lambda_2}4T^2+\frac{\lambda_3}6T^2+\frac{\lambda_4}{12}T^2+\frac{g^2_2}8T^2+\frac{(g_Y^2+g_2^2)}{16}T^2+\frac{|\rho_{tt}|^2}4T^2,
\\
\Pi_{13}={}&\Pi_{31}=\Pi_{24}=\Pi_{42}=0,
\\
\Pi_{12}={}&\Pi_{21}=\Pi_{34}=\Pi_{43}=\frac{\lambda_{7}}4T^2+\frac{y_t\rho_{ttR}}4T^2,
\\
\Pi_{14}={}&\Pi_{41}=-\Pi_{23}=-\Pi_{32}=-\frac{y_t\rho_{ttI}}4T^2,
\\
\Pi_{11}^{charged}={}&\frac{\lambda_1}4T^2+\frac{\lambda_3}6T^2+\frac{\lambda_4}{12}T^2+\frac{g^2_2}8T^2+\frac{(g_Y^2+g_2^2)}{16}T^2+\frac{y^2_t}4T^2,
\\
\Pi_{22}^{charged}={}&\frac{\lambda_2}4T^2+\frac{\lambda_3}6T^2+\frac{\lambda_4}{12}T^2+\frac{g^2_2}8T^2+\frac{(g_Y^2+g_2^2)}{16}T^2+\frac{|\rho_{tt}|^2}4T^2,
\\
\Pi_{12}^{charged}={}&\Pi_{21}^{\ast charged}=\frac{\lambda_{7}}4T^2+\frac{y_t\rho_{tt}}4T^2
\end{align}


For a field-dependent mass squared $x$, we define the contribution
to the Coleman--Weinberg potential per degree of freedom as
\begin{align}
    \mathcal{C}(x;c)
    \equiv
    \frac{x^2}{64\pi^2}
    \left[
        \ln\left(
            \frac{x}{\mu^2}
        \right)
        -
        c
    \right],
\label{eq:CW_compact_function}
\end{align}
where $\mu$ denotes the renormalization scale and
\begin{align}
    c_s
    =
    c_f
    =
    \frac{3}{2},
    \qquad
    c_V
    =
    \frac{5}{6}
\end{align}
for scalar, fermion, and gauge-boson modes, respectively. So, we take $\mu=T$.


In the Arnold--Espinosa prescription\cite{AE}, the unresummed field-dependent
masses are used in the Coleman--Weinberg and ordinary one-loop thermal
contributions.
Only the infrared-sensitive bosonic zero modes are thermally
resummed through a separate ring contribution.
The effective potential is therefore written as
\begin{align}
V_{\mathrm{eff}}^{\mathrm{AE}}(\bm{\phi},T)
={}&
V_0(\bm{\phi})
\nonumber\\
&+
\sum_{a=1}^{4}
\left[
    \mathcal{C}
    \left(
        m_{N,a}^2;\frac{3}{2}
    \right)
    +
    \mathcal{T}_B
    \left(
        m_{N,a}^2;T
    \right)
\right]
\nonumber\\
&+
2\sum_{\alpha=1}^{2}
\left[
    \mathcal{C}
    \left(
        m_{C,\alpha}^2;\frac{3}{2}
    \right)
    +
    \mathcal{T}_B
    \left(
        m_{C,\alpha}^2;T
    \right)
\right]
\nonumber\\
&+
6
\left[
    \mathcal{C}
    \left(
        m_W^2;\frac{5}{6}
    \right)
    +
    \mathcal{T}_B
    \left(
        m_W^2;T
    \right)
\right]
\nonumber\\
&+
3
\left[
    \mathcal{C}
    \left(
        m_Z^2;\frac{5}{6}
    \right)
    +
    \mathcal{T}_B
    \left(
        m_Z^2;T
    \right)
\right]
\nonumber\\
&-
12\,
\mathcal{C}
\left(
    m_t^2;\frac{3}{2}
\right)
+
12\,
\mathcal{T}_F
\left(
    m_t^2;T
\right)
\nonumber\\
&+
V_{\mathrm{ring}}^{\mathrm{AE}}(\bm{\phi},T).
\label{eq:Veff_AE_TB_TF}
\end{align}
Here, the multiplicities $6$, $3$, and $12$ correspond to the
$W$ bosons, the $Z$ boson, and the top quark, respectively.

The ring contribution is
\begin{align}
V_{\mathrm{ring}}^{\mathrm{AE}}(\bm{\phi},T)
={}&
-\frac{T}{12\pi}
\Bigg\{
\sum_{a=1}^{4}
\left[
    \left(
        \overline{m}_{N,a}^2
    \right)^{3/2}
    -
    \left(
        m_{N,a}^2
    \right)^{3/2}
\right]
\nonumber\\
&\quad+
2\sum_{\alpha=1}^{2}
\left[
    \left(
        \overline{m}_{C,\alpha}^2
    \right)^{3/2}
    -
    \left(
        m_{C,\alpha}^2
    \right)^{3/2}
\right]
\nonumber\\
&\quad+
2
\left[
    \left(
        M_{W_L}^2
    \right)^{3/2}
    -
    \left(
        m_W^2
    \right)^{3/2}
\right]
\nonumber\\
&\quad+
\left[
    \left(
        M_{Z_L}^2
    \right)^{3/2}
    -
    \left(
        m_Z^2
    \right)^{3/2}
\right]
+
\left(
    M_{\gamma_L}^2
\right)^{3/2}
\Bigg\}
\label{eq:ring_AE_TB_TF}
\end{align}
where we define thermal resummation
\begin{align}
 \overline m_{N,a}^2(\bm{\phi},T)
 &\equiv
 \operatorname{eig}_a\!\left[
 \mathcal{M}_N^2(\bm{\phi})+\Pi_N(T)
 \right],
 \qquad a=1,\ldots,4,
 \\
 \overline m_{C,\alpha}^2(\bm{\phi},T)
 &\equiv
 \operatorname{eig}_\alpha\!\left[
 \mathcal{M}_C^2(\bm{\phi})+\Pi_C(T)
 \right],
 \qquad \alpha=1,2
\\
M_{W_L}^2(\bm{\phi},T)
    ={}&m_W^2(\bm{\phi})+2g_2^2T^2,
\\
M_{Z_L}^2(\bm{\phi},T)
={}&\frac{1}{2}\left[
(g_2^2+g_Y^2)\left(\frac{\varphi^2}{4}+2T^2\right)
+\Delta_L
\right],
\\
M_{\gamma_L}^2(\bm{\phi},T)
={}&\frac{1}{2}\left[
(g_2^2+g_Y^2)\left(\frac{\varphi^2}{4}+2T^2\right)
-\Delta_L
\right]
\end{align}
as thermal resummation's scalar eigenvalues.
Here, $\Delta_L$ is
\begin{align}
\Delta_L
=\sqrt{
(g_2^2-g_Y^2)^2\left(\frac{\varphi^2}{4}+2T^2\right)^2
+\frac{g_2^2g_Y^2}{4}\varphi^4
}.
\end{align}
The unresummed transverse photon is massless and field independent,
and hence does not contribute after the field-independent thermal
constant has been subtracted.


In the Parwani prescription\cite{Par}, the thermally resummed scalar masses are
inserted directly into both the bosonic Coleman--Weinberg and thermal
functions.
In the gauge sector, we distinguish the transverse and longitudinal
polarizations: the transverse modes are evaluated with their
unresummed field-dependent masses, whereas the Debye-resummed masses
are used for the longitudinal modes.
The effective potential is
\begin{align}
V_{\mathrm{eff}}^{\mathrm{P}}(\bm{\phi},T)
={}&
V_0(\bm{\phi})
\nonumber\\
&+
\sum_{a=1}^{4}
\left[
    \mathcal{C}
    \left(
        \overline{m}_{N,a}^2;\frac{3}{2}
    \right)
    +
    \mathcal{T}_B
    \left(
        \overline{m}_{N,a}^2;T
    \right)
\right]
\nonumber\\
&+
2\sum_{\alpha=1}^{2}
\left[
    \mathcal{C}
    \left(
        \overline{m}_{C,\alpha}^2;\frac{3}{2}
    \right)
    +
    \mathcal{T}_B
    \left(
        \overline{m}_{C,\alpha}^2;T
    \right)
\right]
\nonumber\\
&+
4
\left[
    \mathcal{C}
    \left(
        m_W^2;\frac{5}{6}
    \right)
    +
    \mathcal{T}_B
    \left(
        m_W^2;T
    \right)
\right]
\nonumber\\
&+
2
\left[
    \mathcal{C}
    \left(
        M_{W_L}^2;\frac{5}{6}
    \right)
    +
    \mathcal{T}_B
    \left(
        M_{W_L}^2;T
    \right)
\right]
\nonumber\\
&+
2
\left[
    \mathcal{C}
    \left(
        m_Z^2;\frac{5}{6}
    \right)
    +
    \mathcal{T}_B
    \left(
        m_Z^2;T
    \right)
\right]
\nonumber\\
&+
\left[
    \mathcal{C}
    \left(
        M_{Z_L}^2;\frac{5}{6}
    \right)
    +
    \mathcal{T}_B
    \left(
        M_{Z_L}^2;T
    \right)
\right]
\nonumber\\
&+
\left[
    \mathcal{C}
    \left(
        M_{\gamma_L}^2;\frac{5}{6}
    \right)
    +
    \mathcal{T}_B
    \left(
        M_{\gamma_L}^2;T
    \right)
\right]
\nonumber\\
&-
12\,
\mathcal{C}
\left(
    m_t^2;\frac{3}{2}
\right)
+
12\,
\mathcal{T}_F
\left(
    m_t^2;T
\right).
\label{eq:Veff_Parwani_TB_TF}
\end{align}
The factors $4$ and $2$ in the $W$-boson contributions correspond to
the transverse and longitudinal polarizations, respectively.
Likewise, the transverse and longitudinal $Z$-boson modes have
multiplicities $2$ and $1$.
The longitudinal photon has one degree of freedom.
No additional ring term is included in the Parwani prescription,
since the thermally resummed masses have already been inserted
directly into the bosonic one-loop functions.

\section{Transport Equations and The CP-Violating Source}\label{secondappendix}
The bubble-wall profiles $\phi_i(z)$ obtained from the bounce solution
are used as external inputs to the transport calculation.
We work in the wall rest frame and choose the convention
\begin{align}
    z<0:
    \text{ symmetric phase},
    \qquad
    z>0:
    \text{ broken phase}.
\end{align}
The transport system contains the four effective species
\begin{align}
    a\in\{t_L,b_L,t_R,h\},
\end{align}
where $h$ denotes an effective Higgs fluid containing the scalar
degrees of freedom of the two Higgs doublets.

The quasiparticle distribution function satisfies the semiclassical
Boltzmann equation
\begin{align}
    \left(
        v_{g,a}\partial_z
        +
        F_{a,z}\partial_{p_z}
    \right)
    f_a(z,\bm p)
    =
    C_a[f],
\label{eq:boltzmann-transport}
\end{align}
where $v_{g,a}$, $F_{a,z}$, and $C_a[f]$ denote the group velocity,
the semiclassical force, and the collision term, respectively.
The boosted equilibrium distribution in the wall frame is written as
\begin{align}
    f_{0w,a}
    =
    \frac{1}{
        \exp\left[
            \gamma_w(E_a+v_wp_z)/T
        \right]
        +\eta_a
    },
    \qquad
    \gamma_w
    =
    \frac{1}{\sqrt{1-v_w^2}},
\label{eq:boosted-equilibrium-distribution}
\end{align}
where $\eta_a=+1$ for fermions and $\eta_a=-1$ for bosons.

For each particle species, we retain the chemical-potential and
velocity moments,
\begin{align}
    \bm w_a(z)
    =
    \begin{pmatrix}
        \mu_a(z)\\
        u_a(z)
    \end{pmatrix}.
\end{align}
The resulting finite-wall-velocity two-moment equations\cite{tra1} are
\begin{align}
    -D_{1a}\mu_a'
    +u_a'
    +v_w\gamma_w(m_a^2)'Q_{1a}\mu_a
    -C_a
    &=
    S_{1a},
\label{eq:transport-moment-zero}
    \\
    -D_{2a}\mu_a'
    +R_a u_a'
    +v_w\gamma_w(m_a^2)'Q_{2a}\mu_a
    +(m_a^2)'\overline{R}_a u_a
    +\Gamma_{{\rm tot},a}u_a
    +v_w C_a
    &=
    S_{2a},
\label{eq:transport-moment-one}
\end{align}
where a prime denotes differentiation with respect to $z$ and
\begin{align}
    R_a=-v_w.
\end{align}
The quantities $\Gamma_{{\rm tot},a}$ describe the collisional
relaxation of the velocity moments.

In Section \ref{sec3}, we take $0<v_\omega\le0.6$.
\subsection{Finite-Wall-Velocity Transport Coefficients}

We introduce the dimensionless variables
\begin{align}
    q
    &=
    \frac{|\bm p|}{T},
    &
    y
    &=
    \frac{p_z}{|\bm p|},
    &
    x_a
    &=
    \frac{m_a}{T},
    \\
    \epsilon_a
    &=
    \sqrt{q^2+x_a^2},
    &
    \epsilon_{z,a}
    &=
    \sqrt{q^2y^2+x_a^2},
    &
    \xi_a
    &=
    \gamma_w(\epsilon_a+v_wqy).
\end{align}
We further define
\begin{align}
    f_a(\xi_a)
    &=
    \frac{1}{e^{\xi_a}+\eta_a},
    &
    \mathrm{d}\Omega_q
    &=
    2\pi q^2\,\mathrm{d}q\,\mathrm{d}y,
    \\
    N_{1a}
    &=
    \gamma_w\widehat N_{1a},
    &
    \widehat N_{1a}
    &=
    -\frac{2\pi^3}{3}.
\end{align}
Derivatives of $f_a$ below are taken with respect to $\xi_a$.
The coefficients entering Eqs.~\eqref{eq:transport-moment-zero} and
\eqref{eq:transport-moment-one} are evaluated as
\begin{align}
    D_{0a}
    &=
    \frac{1}{\widehat N_{1a}}
    \int_0^\infty
    4\pi q^2\,\mathrm{d}q\,
    f_{0,a}'(\epsilon_a),
    \\
    D_{1a}
    &=
    -v_wD_{0a},
    \\
    D_{2a}
    &=
    \frac{1}{N_{1a}}
    \int\mathrm{d}\Omega_q\,
    \left(
        \frac{qy}{\epsilon_a}
    \right)^2
    f_a'(\xi_a),
    \\
    Q_{1a}
    &=
    \frac{1}{T^2N_{1a}}
    \int\mathrm{d}\Omega_q\,
    \frac{f_a''(\xi_a)}{2\epsilon_a},
    \\
    Q_{2a}
    &=
    \frac{1}{T^2N_{1a}}
    \int\mathrm{d}\Omega_q\,
    \frac{qy\,f_a''(\xi_a)}{2\epsilon_a^2},
    \\
    K_{0a}
    &=
    -\frac{T}{N_{1a}}
    \int\mathrm{d}\Omega_q\,
    f_a(\xi_a).
\label{eq:finite-vw-coefficients}
\end{align}
The coefficient $\overline R_a$ is evaluated from the corresponding integral,
\begin{align}
    \overline R_a
    =
    \frac{\pi}{
        \gamma_w^2N_{0a}T^2
    }
    \int_0^\infty \mathrm{d}q\,
    \frac{q}{\epsilon_a}
    \ln\left|
        \frac{q-v_w\epsilon_a}{
              q+v_w\epsilon_a}
    \right|
    f_{0,a}(\epsilon_a),
\label{eq:Rbar-coefficient}
\end{align}
where $N_{0a}$ denotes the corresponding zeroth-moment normalization.
In the numerical calculation, these coefficients are tabulated as
functions of $m_a/T$ and $v_w$ and interpolated along the bubble wall.

\subsection{Explicit CP-Violating Source}

The field-dependent complex top-quark mass along the wall is
\begin{align}
    m_t(z)
    &=
    |m_t(z)|e^{i\theta_t(z)}
    =
    \frac{A(z)+iB(z)}{\sqrt{2}}
    \nonumber\\
    &=
    \frac{1}{\sqrt{2}}
    \left[
        y_t\phi_1(z)
        +
        \rho_{tt}
        \left(
            \phi_2(z)+i\phi_3(z)
        \right)
    \right],
\label{eq:complex-top-mass-wall}
\end{align}
where
\begin{align}
    \rho_{tt}
    &=
    \rho_{ttR}+i\rho_{ttI},
    \\
    A(z)
    &=
    y_t\phi_1(z)
    +\rho_{ttR}\phi_2(z)
    -\rho_{ttI}\phi_3(z),
    \\
    B(z)
    &=
    \rho_{ttR}\phi_3(z)
    +\rho_{ttI}\phi_2(z).
\end{align}
Instead of numerically differentiating the phase $\theta_t(z)$, we
use the identity
\begin{align}
    q_t(z)
    \equiv
    |m_t(z)|^2\theta_t'(z)
    =
    \frac{1}{2}
    \left[
        A(z)B'(z)-B(z)A'(z)
    \right].
\label{eq:m2-theta-prime}
\end{align}
The two CP-odd source moments are
\begin{align}
    S_\ell(z)
    =
    \gamma_wv_w
    \left[
        -Q_\ell^8(z)q_t'(z)
        +
        Q_\ell^9(z)q_t(z)
        \bigl(m_t^2(z)\bigr)'
    \right],
    \qquad
    \ell=1,2.
\label{eq:CPV-source-moments}
\end{align}
The thermal source coefficients are
\begin{align}
    Q_1^8
    &=
    \frac{1}{T^2N_{1t}}
    \int\mathrm{d}\Omega_q\,
    \frac{
        \operatorname{sgn}(qy)f_t'(\xi_t)
    }{
        2\epsilon_t\epsilon_{z,t}
    },
    \\
    Q_2^8
    &=
    \frac{1}{T^2N_{1t}}
    \int\mathrm{d}\Omega_q\,
    \frac{
        |qy|f_t'(\xi_t)
    }{
        2\epsilon_t^2\epsilon_{z,t}
    },
    \\
    Q_1^9
    &=
    \frac{1}{T^4N_{1t}}
    \int\mathrm{d}\Omega_q\,
    \frac{
        \operatorname{sgn}(qy)
    }{
        4\epsilon_t^2\epsilon_{z,t}
    }
    \left[
        \frac{f_t'(\xi_t)}{\epsilon_t}
        -
        \gamma_w f_t''(\xi_t)
    \right],
    \\
    Q_2^9
    &=
    \frac{1}{T^4N_{1t}}
    \int\mathrm{d}\Omega_q\,
    \frac{
        |qy|
    }{
        4\epsilon_t^3\epsilon_{z,t}
    }
    \left[
        \frac{f_t'(\xi_t)}{\epsilon_t}
        -
        \gamma_w f_t''(\xi_t)
    \right].
\label{eq:CPV-source-coefficients}
\end{align}
In our particle-species convention, the sources are assigned as
\begin{align}
    S_{\ell,t_L}
    &=
    +S_\ell,
    &
    S_{\ell,t_R}
    &=
    -S_\ell,
    &
    S_{\ell,b_L}
    &=
    S_{\ell,h}
    =
    0.
\label{eq:source-assignment}
\end{align}
The overall sign depends on the wall-coordinate and particle--antiparticle
conventions, whereas the opposite signs for $t_L$ and $t_R$ encode
the chiral structure of the source.

\subsection{Collision Terms}

We define the combinations
\begin{align}
    \Delta_t
    &=
    \mu_{t_L}-\mu_{t_R}+\mu_h,
    &
    \Delta_b
    &=
    \mu_{b_L}-\mu_{t_R}+\mu_h,
    \\
    \Delta_M
    &=
    \mu_{t_L}-\mu_{t_R},
    &
    \Delta_W
    &=
    \mu_{t_L}-\mu_{b_L},
\end{align}
and
\begin{align}
    \Delta_{\rm ss}
    ={}&
    \left(1+9D_{0t}\right)\mu_{t_L}
    +
    \left(1+9D_{0b}\right)\mu_{b_L}
    \nonumber\\
    &-
    \left(1-9D_{0t}\right)\mu_{t_R}.
\end{align}
The reaction terms entering the collision network are
\begin{align}
    \mathcal R_{t_L}
    ={}&
    \Gamma_y\Delta_t
    +
    \Gamma_M\Delta_M
    +
    \Gamma_W\Delta_W
    +
    \Gamma_{\rm ss}\Delta_{\rm ss},
    \\
    \mathcal R_{b_L}
    ={}&
    \Gamma_y\Delta_b
    -
    \Gamma_W\Delta_W
    +
    \Gamma_{\rm ss}\Delta_{\rm ss},
    \\
    \mathcal R_{t_R}
    ={}&
    -\Gamma_y(\Delta_t+\Delta_b)
    -
    \Gamma_M\Delta_M
    -
    \Gamma_{\rm ss}\Delta_{\rm ss},
    \\
    \mathcal R_h
    ={}&
    \Gamma_y(\Delta_t+\Delta_b)
    +
    \Gamma_h\mu_h.
\label{eq:reaction-network}
\end{align}
The collision quantities in the transport equations are
\begin{align}
    C_a(z)
    =
    \frac{K_{0a}(z)}{T}
    \mathcal R_a(z).
\label{eq:collision-quantity}
\end{align}
Here, $\Gamma_y$, $\Gamma_M$, $\Gamma_W$,
$\Gamma_{\rm ss}$, and $\Gamma_h$ describe the top-Yukawa interaction,
top chirality relaxation, $t_L\leftrightarrow b_L$ exchange, the strong
sphaleron, and Higgs-number relaxation, respectively.

The benchmark rates\cite{tra2} are parametrized as
\begin{align}
    \Gamma_y
    &=
    c_yT,
    &
    \Gamma_{\rm ss}
    &=
    c_{\rm ss}T,
    &
    \Gamma_M(z)
    &=
    \frac{m_t^2(z)}{63T},
    &
    \Gamma_h(z)
    &=
    \frac{m_W^2(z)}{50T}.
\label{eq:benchmark-reaction-rates}
\end{align}
Now, $c_y$ and $c_{\rm ss}$ are $4.2\times10^{-4}$ and $4.9\times10^{-4}$. So, $\Gamma_W$ is identified with the total Higgs-fluid relaxation rate.

The representative diffusion constants\cite{tra2} are
\begin{align}
    D_t
    =
    D_b
    =
    \frac{6}{T},
    \qquad
    D_h
    =
    \frac{20}{T}.
\label{eq:diffusion-constants}
\end{align}

The neutral and charged scalar modes of the two Higgs doublets are
combined into one effective Higgs fluid. For a scalar mode $\alpha$
with real multiplicity $g_\alpha$, a generic thermal coefficient is
averaged according to
\begin{align}
    \langle Q_{n,\alpha}\rangle_w
    &=
    \frac{
        \sum_\alpha
        g_\alpha Q_{n,\alpha}
    }{
        \sum_\alpha g_\alpha
    },
    \\
    \left\langle
        (m_\alpha^2)'Q_{n,\alpha}
    \right\rangle_w
    &=
    \frac{
        \sum_\alpha
        g_\alpha
        (m_\alpha^2)'Q_{n,\alpha}
    }{
        \sum_\alpha g_\alpha
    }.
\label{eq:Higgs-fluid-average}
\end{align}
In particular, the mass-gradient contribution is averaged after
forming the product $(m_\alpha^2)'Q_{n,\alpha}$ for each mode.

\section{Baryon-Number Evolution and The Baryon Asymmetry}\label{thirdappendix} 
In the linear-response approximation, the particle densities are
related to the chemical potentials through
\begin{align}
    n_a(z)
    =
    \frac{k_aT^2}{6}\mu_a(z),
\label{eq:density-chemical-potential}
\end{align}
where the effective statistical weights are
\begin{align}
    \left(
        k_{t_L},
        k_{b_L},
        k_{t_R},
        k_h
    \right)
    =
    (3,3,3,8).
\end{align}
The value $k_h=8$ represents the eight real scalar degrees of freedom
of the two Higgs doublets combined into the effective Higgs fluid. In principle, the finite thermal masses of the scalar quasiparticles,
which are not necessarily negligible in the parameter region with
$\lambda_3>1$, modify the Higgs-sector susceptibility and shift
$k_h$ away from its relativistic value, $k_h=8$.
For simplicity, we neglect this correction in the
density--chemical-potential conversion and set $k_h=8$ throughout
the present analysis.

The left-handed baryon chemical potential that biases the weak
sphaleron\cite{tra1} is
\begin{align}
    \mu_{B_L}(z)
    ={}&
    \frac{1}{2}
    \left[
        1+4D_{0t}(z)
    \right]
    \mu_{t_L}(z)
    \nonumber\\
    &+
    \frac{1}{2}
    \left[
        1+4D_{0b}(z)
    \right]
    \mu_{b_L}(z)
    +
    2D_{0t}(z)\mu_{t_R}(z).
\label{eq:left-handed-baryon-potential}
\end{align}
The positive sign of the last term follows from our convention in
which the transported species is the right-handed top $t_R$, rather
than the charge-conjugated field $t^c$.

We denote the weak-sphaleron transition rate per unit volume by
$\Gamma_{\rm ws}^{(4)}$ and define the dimension-one rate
\begin{align}
    \overline\Gamma_{\rm ws}
    \equiv
    \frac{\Gamma_{\rm ws}^{(4)}}{T^3}.
\label{eq:weak-sphaleron-rate}
\end{align}
The local baryon-number density then obeys
\begin{align}
    \frac{\mathrm{d}n_B}{\mathrm{d}t}
    =
    \frac{3}{2}
    \overline\Gamma_{\rm ws}
    \left[
        3T^2\mu_{B_L}(t)f_{\rm sph}(t)
        -
        \frac{15}{2}n_B(t)
    \right].
\label{eq:baryon-time-evolution}
\end{align}
Equivalently,
\begin{align}
    \frac{\mathrm{d}n_B}{\mathrm{d}t}
    =
    \frac{9}{2}
    \overline\Gamma_{\rm ws}
    T^2\mu_{B_L}f_{\rm sph}
    -
    \frac{45}{4}
    \overline\Gamma_{\rm ws}n_B,
\end{align}
where the first and second terms describe baryon-number generation
and washout, respectively. The Hubble-expansion term is neglected
because the local wall-crossing time is much shorter than the Hubble
time.

On the symmetric-phase side, we introduce
\begin{align}
    x=-z\geq0,
    \qquad
    v_g=\gamma_wv_w.
\end{align}
Since $\mathrm{d}x/\mathrm{d}t=-v_g$, Eq.~\eqref{eq:baryon-time-evolution}
becomes
\begin{align}
    -v_g\frac{\mathrm{d}n_B}{\mathrm{d}x}
    =
    \frac{9}{2}
    \overline\Gamma_{\rm ws}
    T^2\mu_{B_L}(x)f_{\rm sph}(x)
    -
    \frac{45}{4}
    \overline\Gamma_{\rm ws}n_B(x).
\label{eq:baryon-spatial-evolution}
\end{align}
Imposing
\begin{align}
    n_B(x\to\infty)=0,
\end{align}
the baryon-number density captured by the wall is
\begin{align}
    n_B(0)
    =
    \frac{
        9\overline\Gamma_{\rm ws}T^2
    }{
        2v_g
    }
    \int_0^\infty \mathrm{d}x\,
    \mu_{B_L}(x)f_{\rm sph}(x)
    \exp\left[
        -\frac{
            45\overline\Gamma_{\rm ws}
        }{
            4v_g
        }x
    \right].
\label{eq:baryon-density-integral}
\end{align}
Using the entropy density
\begin{align}
    s
    =
    \frac{2\pi^2}{45}g_*T^3,
\end{align}
the baryon-to-entropy ratio is
\begin{align}
    Y_B
    \equiv
    \frac{n_B}{s}
    =
    \frac{
        405\overline\Gamma_{\rm ws}
    }{
        4\pi^2g_*T\gamma_wv_w
    }
    \int_0^\infty \mathrm{d}x\,
    \mu_{B_L}(x)f_{\rm sph}(x)
    \exp\left[
        -\frac{
            45\overline\Gamma_{\rm ws}
        }{
            4\gamma_wv_w
        }x
    \right].
\label{eq:final-baryon-asymmetry}
\end{align}

The local electroweak order parameter is defined as
\begin{align}
    v(z)
    =
    \sqrt{
        \phi_1^2(z)
        +
        \phi_2^2(z)
        +
        \phi_3^2(z)
    }.
\end{align}
The phenomenological sphaleron-suppression profile used in the
numerical calculation\cite{tra1} is
\begin{align}
    f_{\rm sph}(z)
    =
    \min\left[
        1,\,
        \kappa_{\rm sph}
        \frac{T}{
            \overline\Gamma_{\rm ws}
        }
        \exp\left(
            -c_{\rm sph}\frac{v(z)}{T}
        \right)
    \right],
\label{eq:sphaleron-suppression-profile}
\end{align}
with
\begin{align}
    \kappa_{\rm sph}
    =
    2.4,
    \qquad
    c_{\rm sph}
    =
    40,
    \qquad
    \overline\Gamma_{\rm ws}=10^{-6}T.
\end{align}
Thus, $f_{\rm sph}\simeq1$ in the symmetric phase and becomes
exponentially suppressed toward the broken phase.

\section{Electric and Chromoelectric Dipole Moments}\label{fourthappendix}

The CP-violating interactions required for electroweak baryogenesis
generically induce electric and chromoelectric dipole moments.
In the present analysis, the additional neutral scalars first generate
top-quark dipole operators at the heavy-scalar scale. These operators
are subsequently evolved to lower energies through renormalization-group
running and heavy-particle threshold matching, thereby inducing the
electric dipole moments of the electron, neutron, and proton:
\begin{align}
    \text{heavy-scalar matching}
    \longrightarrow
    \left\{
        d_t,\widetilde d_t,
        C_{tB},C_{tW}
    \right\}
    \longrightarrow
    \left\{
        d_e,d_n,d_p
    \right\}.
\end{align}

We define the relevant CP-odd effective operators by
\begin{align}
\mathcal{L}_{\mathrm{CPV}}
={}&
-\frac{1}{2}d_\psi\,
\overline{\psi}\sigma_{\mu\nu}i\gamma^5\psi
F^{\mu\nu}
\nonumber\\
&-
\frac{1}{2}g_s\widetilde d_q\,
\overline q\sigma_{\mu\nu}i\gamma^5T^a q
G_a^{\mu\nu}
\nonumber\\
&+
\frac{1}{3}w f_{abc}
G_{\mu\nu}^a
\widetilde G^{b\,\nu\rho}
G^c{}_{\rho}{}^{\mu},
\label{eq:cp_odd_effective_operators}
\end{align}
where $d_\psi$ denotes a fermion EDM, $\widetilde d_q$ denotes a
quark CEDM, and $w$ is the coefficient of the CP-odd Weinberg
three-gluon operator. We use
\begin{align}
    \widetilde G^{a\,\mu\nu}
    \equiv
    \frac{1}{2}
    \epsilon^{\mu\nu\rho\sigma}
    G^a_{\rho\sigma}.
\end{align}
The mass dimensions of the Wilson coefficients are
\begin{align}
    [d_\psi]
    =
    [\widetilde d_q]
    =
    \mathrm{GeV}^{-1},
    \qquad
    [w]
    =
    \mathrm{GeV}^{-2}.
\end{align}
For conversion to conventional EDM units, we use
\begin{align}
    1~\mathrm{GeV}^{-1}
    =
    1.973269804\times10^{-14}~\mathrm{cm}.
\end{align}

For comparison with \cite{edm1,edm2,Sahoo2017}, 
\begin{align}
    |d_e|
    &<
    4.1\times10^{-30}\ e\,\mathrm{cm},
    \label{eq:electron_edm_limit}
    \\
    |d_n|
    &<
    1.0\times10^{-26}\ e\,\mathrm{cm},
    \label{eq:neutron_edm_limit}
    \\
    |d_p|
    &<
    2.1\times10^{-25}\ e\,\mathrm{cm}.
    \label{eq:proton_edm_limit}
\end{align}

\subsection{One-loop Top-Quark Dipole Moments}
\label{subsec:one_loop_top_dipoles}

Let $H_i$ denote the three physical neutral scalar mass eigenstates,
and let $\mathcal R_{ai}$ be the real orthogonal matrix that rotates
the neutral real-field basis into the mass basis. The general
one-loop top CEDM is
\begin{align}
\widetilde d_t^{(1)}
={}&
-\frac{m_t}{16\pi^2}
\sum_{i=1}^{3}
\Bigl(
    y_t\mathcal R_{1i}
    +
    \rho_{ttR}\mathcal R_{2i}
    +
    \rho_{ttI}\mathcal R_{3i}
\Bigr)
\nonumber\\
&\hspace{18mm}\times
\Bigl(
    \rho_{ttI}\mathcal R_{2i}
    -
    \rho_{ttR}\mathcal R_{3i}
\Bigr)
C_{11}[H_i,t,t].
\label{eq:general_one_loop_top_cedm}
\end{align}
The loop function is defined by
\begin{align}
    C_{11}(m_\varphi^2)
    \equiv
    C_{11}[m_\varphi^2;t,t]
    =
    \int_0^1\mathrm{d}x\,
    \frac{x^2}{
        m_t^2x^2
        -
        m_\varphi^2x
        +
        m_\varphi^2
    }.
\label{eq:C11_loop_function}
\end{align}
Thus,
\begin{align}
    [C_{11}]
    =
    \mathrm{GeV}^{-2},
    \qquad
    [\widetilde d_t]
    =
    \mathrm{GeV}^{-1}.
\end{align}

\begin{figure*}[t]
    \centering
    \includegraphics[width=0.88\textwidth]{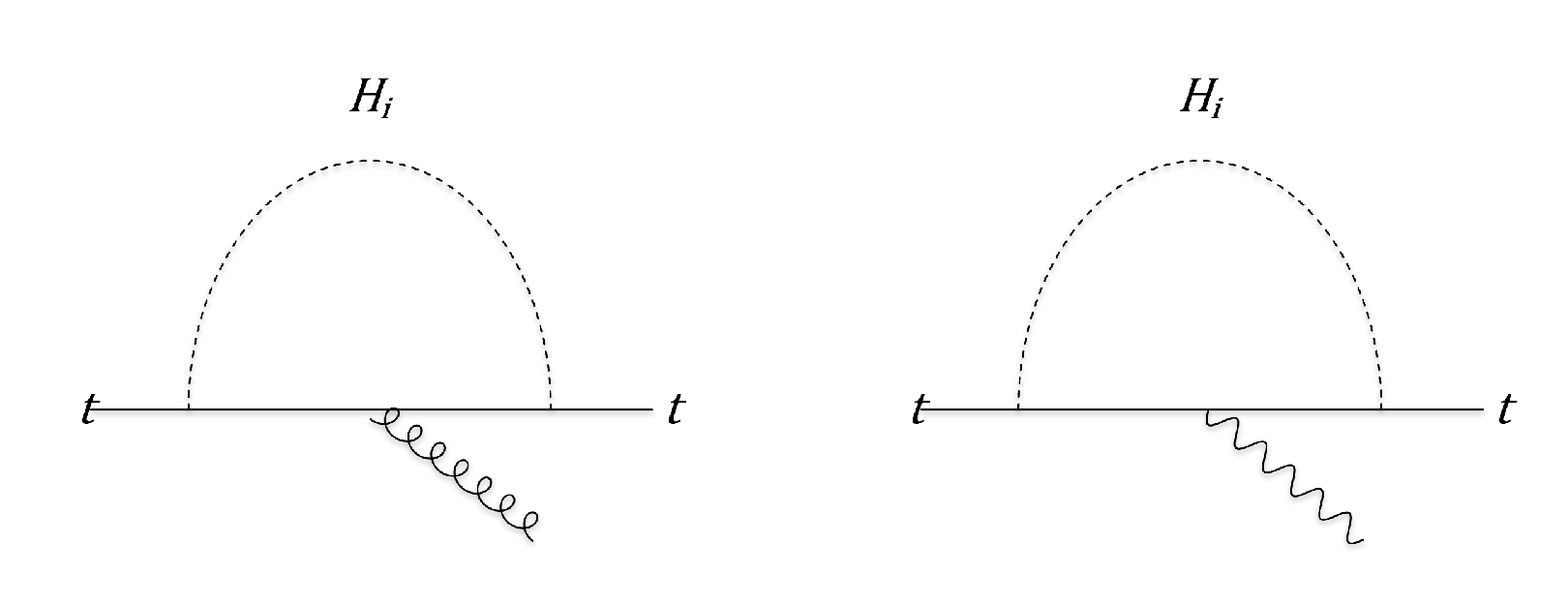}
    \caption{
    Representative one-loop neutral-scalar contributions to the
    top-quark CEDM (left) and EDM (right).
    The dashed line denotes a neutral scalar mass eigenstate $H_i$,
    while the external curly and wavy lines denote a gluon and a photon,
    respectively.
    }
    \label{fig:one_loop_top_dipoles}
\end{figure*}

In the alignment limit with $\lambda_6=0$ and negligible neutral-scalar
mixing, the nonstandard neutral scalars can be identified with the
CP-even state $H$ and the CP-odd state $A$. Writing
\begin{align}
    \rho_{tt}
    =
    \rho_{ttR}
    +
    i\rho_{ttI},
\end{align}
Eq.~\eqref{eq:general_one_loop_top_cedm} reduces to
\begin{align}
    \widetilde d_t^{(1)}
    =
    -\frac{m_t}{16\pi^2}
    \rho_{ttR}\rho_{ttI}
    \left[
        C_{11}(m_H^2)
        -
        C_{11}(m_A^2)
    \right].
\label{eq:one_loop_top_cedm_alignment}
\end{align}
The corresponding top EDM is obtained from the same charge-stripped
loop coefficient:
\begin{align}
    \frac{d_t^{(1)}}{e}
    =
    Q_t\widetilde d_t^{(1)},
    \qquad
    Q_t=\frac{2}{3}.
\label{eq:one_loop_top_edm_alignment}
\end{align}
The representative one-loop diagrams contributing to the top-quark
CEDM and EDM are shown in Fig.~\ref{fig:one_loop_top_dipoles}.
The two contributions are described by the same charge-stripped loop
coefficient, with the external gluon replaced by a photon for the EDM.

It is useful to express the result in terms of the rephasing-invariant
CP-odd combination
\begin{align}
    J_5
    \equiv
    \operatorname{Im}
    \left(
        \lambda_5\rho_{tt}^2
    \right).
\label{eq:J5_invariant}
\end{align}
For real $\lambda_5$,
\begin{align}
    J_5
    =
    2\lambda_5
    \rho_{ttR}\rho_{ttI}.
\end{align}
Using the mass relation
\begin{align}
    m_H^2-m_A^2
    =
    \lambda_5v^2,
\end{align}
we obtain
\begin{align}
    \rho_{ttR}\rho_{ttI}
    =
    \frac{
        v^2J_5
    }{
        2(m_H^2-m_A^2)
    },
\end{align}
and hence
\begin{align}
    \boxed{
    \widetilde d_t^{(1)}
    =
    -\frac{m_tv^2}{32\pi^2}
    J_5
    \frac{
        C_{11}(m_H^2)-C_{11}(m_A^2)
    }{
        m_H^2-m_A^2
    }
    }.
\label{eq:one_loop_top_cedm_J5}
\end{align}

When $\lambda_5=0$, one has
\begin{align}
    m_H=m_A,
    \qquad
    J_5=0,
\end{align}
and the $H$ and $A$ contributions cancel:
\begin{align}
    \widetilde d_t^{(1)}
    =
    d_t^{(1)}
    =
    0.
\label{eq:one_loop_dipole_degenerate_zero}
\end{align}
This cancellation is a consequence of the degeneracy of the CP-even
and CP-odd contributions and is not merely due to a small coupling.
For nonzero $\lambda_6$, an additional invariant proportional to
$\operatorname{Im}(\lambda_6\rho_{tt})$ can generate a one-loop
dipole through neutral-scalar mixing.

A charged-Higgs one-loop contribution would require two independent
chiral couplings,
\begin{align}
    \mathcal L
    \supset
    H^+\overline t
    \left(
        g_LP_L+g_RP_R
    \right)b
    +
    \mathrm{h.c.},
\end{align}
and is proportional to
$\operatorname{Im}(g_Lg_R^\ast)$. In the top-philic limit
used here,
\begin{align}
    \rho_{bb}
    =
    \rho_{tc}
    =
    \rho_{ct}
    =
    \cdots
    =
    0,
\end{align}
so that no independent second chiral nonstandard coupling is present.
We therefore set
\begin{align}
    d_t^{(1),H^\pm}
    =
    \widetilde d_t^{(1),H^\pm}
    =
    0.
\end{align}

\subsection{Two-loop Top CEDM}
\label{subsec:two_loop_top_cedm}

The leading scalar-self-energy contribution is controlled by the
second rephasing-invariant CP-odd combination
\begin{align}
    J_7
    \equiv
    \operatorname{Im}
    \left(
        \lambda_7\rho_{tt}
    \right).
\label{eq:J7_invariant}
\end{align}
In the minimal limit
\begin{align}
    \lambda_5
    =
    \lambda_6
    =
    0,
\end{align}
the one-loop dipole vanishes, while a nonzero $J_7$ generates the
leading top EDM and CEDM at two loops\cite{edm3}.

\begin{figure*}[t]
    \centering
    \includegraphics[width=0.97\textwidth]{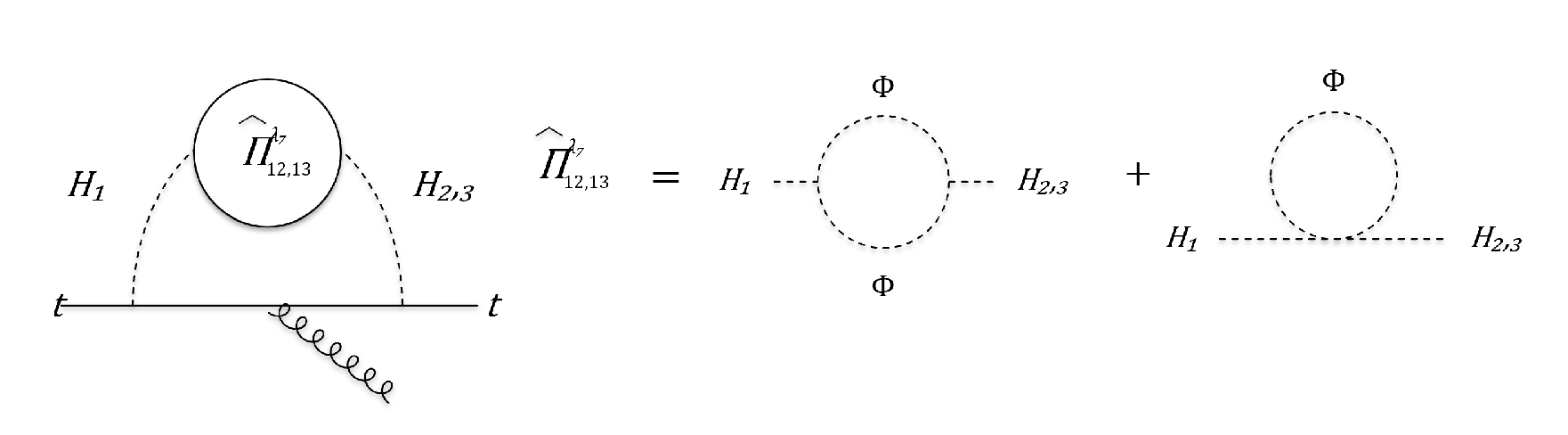}
    \caption{
    Leading type-A two-loop contribution to the top-quark CEDM.
    The left panel shows the insertion of the renormalized off-diagonal
    scalar self-energy
    $\widehat{\Pi}_{12,13}^{\lambda_7}$
    into the neutral-scalar top dipole loop.
    The right panel shows the scalar-loop contributions to
    $\widehat{\Pi}_{12,13}^{\lambda_7}$, where $\Phi$ denotes the scalar
    states propagating inside the self-energy loop.
    }
    \label{fig:two_loop_typeA_top_cedm}
\end{figure*}

The diagrammatic origin of the leading type-A contribution is
illustrated in Fig.~\ref{fig:two_loop_typeA_top_cedm}.
It can be understood as the insertion of a $\lambda_7$-induced
renormalized off-diagonal scalar self-energy into the neutral-scalar
top dipole loop.

For degenerate nonstandard scalars,
\begin{align}
    m_H
    =
    m_A
    =
    m_{H^\pm}
    \equiv
    m_\Phi,
\end{align}
the leading type-A contribution is
\begin{equation}
\begin{aligned}
\widetilde d_t^{(2),A}
={}&
\frac{
    \operatorname{Im}
    (\lambda_7\rho_{tt})
}{
    \sqrt{2}
}
\frac{
    3\lambda_3v
}{
    (16\pi^2)^2
}
\frac{
    2m_t^2
}{
    m_\Phi^2-m_h^2
}
\nonumber\\
&\times
\int_{4m_\Phi^2}^{\infty}
\mathrm{d}s\,
\frac{
    \lambda^{1/2}
    (s,m_\Phi^2,m_\Phi^2)
}{
    s
}
\mathcal B(s),
\end{aligned}
\label{eq:two_loop_typeA_degenerate}
\end{equation}
where these functions are
\begin{align}
    \lambda(a,b,c)
    =
    (a-b-c)^2-4bc,
\end{align}
and
\begin{align}
\mathcal B(s)
={}&
\frac{
    C_{11}(s)-C_{11}(m_\Phi^2)
}{
    m_\Phi^2-s
}
-
\frac{
    C_{11}(s)-C_{11}(m_h^2)
}{
    m_h^2-s
}
\nonumber\\
&-
\frac{
    C_{11}(m_\Phi^2)-C_{11}(m_h^2)
}{
    s
}.
\label{eq:typeA_B_function}
\end{align}
This contribution can be viewed as a scalar-loop-induced,
renormalized $h$--$H$ or $h$--$A$ self-energy insertion into the
top dipole loop. 

For the nondegenerate scalar spectrum used in the numerical scan,
we employ the following leading extension. We first define
\begin{align}
\mathcal K(m_L,m_E)
={}&
\frac{
    2m_t^2
}{
    m_E^2-m_h^2
}
\int_{4m_L^2}^{\infty}
\mathrm{d}s\,
\frac{
    \lambda^{1/2}(s,m_L^2,m_L^2)
}{
    s
}
\nonumber\\
&\times
\Bigg[
\frac{
    C_{11}(s)-C_{11}(m_E^2)
}{
    m_E^2-s
}
-
\frac{
    C_{11}(s)-C_{11}(m_h^2)
}{
    m_h^2-s
}
\nonumber\\
&\hspace{27mm}
-
\frac{
    C_{11}(m_E^2)-C_{11}(m_h^2)
}{
    s
}
\Bigg].
\label{eq:nondegenerate_typeA_kernel}
\end{align}
For $X=H,A,C$, with
\begin{align}
    m_C
    \equiv
    m_{H^\pm},
\end{align}
the scalar weights are
\begin{align}
    w_H^{HH}
    &=
    \frac{3}{2}
    (\lambda_3+\lambda_4+\lambda_5),
    &
    w_H^{AA}
    &=
    \frac{1}{2}
    (\lambda_3+\lambda_4-\lambda_5),
    \nonumber\\
    w_H^{CC}
    &=
    \lambda_3,
    &
    w_A^{HH}
    &=
    \frac{1}{2}
    (\lambda_3+\lambda_4+\lambda_5),
    \nonumber\\
    w_A^{AA}
    &=
    \frac{3}{2}
    (\lambda_3+\lambda_4-\lambda_5),
    &
    w_A^{CC}
    &=
    \lambda_3.
\label{eq:nondegenerate_typeA_weights}
\end{align}
The nondegenerate contribution used in the scan is then
\begin{equation}
\begin{aligned}
\widetilde d_t^{(2),A,\mathrm{ND}}
={}&
\frac{
    v
}{
    \sqrt{2}(16\pi^2)^2
}
\Bigg[
\lambda_{7R}\rho_{ttI}
\sum_{X=H,A,C}
w_H^{XX}
\mathcal K(m_X,m_H)
\nonumber\\
&\hspace{18mm}
+
\lambda_{7I}\rho_{ttR}
\sum_{X=H,A,C}
w_A^{XX}
\mathcal K(m_X,m_A)
\Bigg].
\end{aligned}
\label{eq:two_loop_typeA_nondegenerate}
\end{equation}
For
\begin{align}
    \lambda_4
    =
    \lambda_5
    =
    0,
    \qquad
    m_H
    =
    m_A
    =
    m_{H^\pm},
\end{align}
each weighted sum reduces to $3\lambda_3$, and
Eq.~\eqref{eq:two_loop_typeA_nondegenerate} reproduces the degenerate
expression in Eq.~\eqref{eq:two_loop_typeA_degenerate}.

\subsection{Top Electroweak Dipoles and The Electron EDM}
\label{subsec:electron_edm}
The electron EDM in the general 2HDM has been studied at the
complete two-loop level in \cite{edm3}.
In the top-philic limit considered here, we instead evaluate the
electron EDM induced through the top electroweak dipole operators\cite{edm4}.

To evaluate the electron EDM, we introduce the top electroweak
dipole operators at the matching scale $\Lambda$:
\begin{align}
\mathcal L_{\mathrm{eff}}
=
-\frac{1}{\Lambda^2}
\Bigg[
&
\frac{g_1}{\sqrt{2}}
C_{tB}
\overline Q_L
\sigma^{\mu\nu}
t_R
\widetilde\Phi_1
B_{\mu\nu}
\nonumber\\
&+
\frac{g_2}{\sqrt{2}}
C_{tW}
\overline Q_L
\sigma^{\mu\nu}
\tau^a t_R
\widetilde\Phi_1
W_{\mu\nu}^a
+
\mathrm{h.c.}
\Bigg].
\label{eq:top_electroweak_dipole_operators}
\end{align}
After electroweak symmetry breaking,
\begin{align}
    \operatorname{Im}C_{tB}
    &=
    \frac{\Lambda^2}{g_1v}
    d_t^B,
    &
    \operatorname{Im}C_{tW}
    &=
    \frac{\Lambda^2}{g_2v}
    d_t^{W^3}.
\label{eq:top_ew_dipole_matching}
\end{align}

Defining the one-loop charge-stripped coefficient by
\begin{align}
    \mathcal K_1
    \equiv
    \widetilde d_t^{(1)},
\end{align}
the one-loop electroweak dipole coefficients are
\begin{align}
    d_t^{B,(1)}
    &=
    \frac{5g_1}{12}
    \mathcal K_1,
    &
    d_t^{W^3,(1)}
    &=
    \frac{g_2}{4}
    \mathcal K_1.
\label{eq:one_loop_top_ew_dipoles}
\end{align}

Similarly, denoting the charge-stripped type-A and type-B two-loop
coefficients by $\mathcal K_A$ and $\mathcal K_B$, respectively,
\begin{align}
    d_t^{B,(2)}
    &=
    \frac{5g_1}{12}\mathcal K_A
    +
    \frac{g_1}{2}\mathcal K_B,
    \\
    d_t^{W^3,(2)}
    &=
    \frac{g_2}{4}\mathcal K_A
    +
    \frac{g_2}{2}\mathcal K_B.
\label{eq:two_loop_top_ew_dipoles}
\end{align}
The type-A coefficient is obtained from the same scalar-self-energy
kernel as the top CEDM. The type-B contribution corresponds to a
different electroweak topology involving additional three-point loop
functions. 

\begin{figure}[t]
    \centering
    \includegraphics[width=\textwidth]{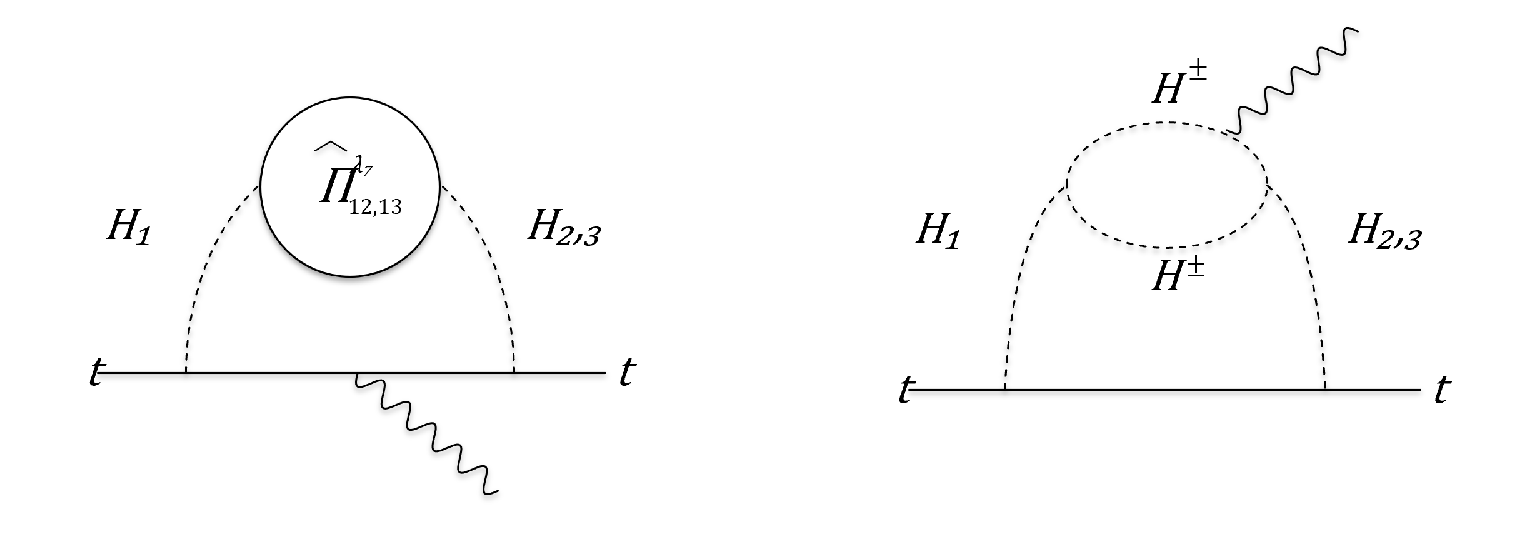}
    \caption{
    Representative two-loop contributions to the top-quark
    electroweak dipole operators.
    The left panel shows the type-A contribution, in which the
    $\lambda_7$-induced renormalized off-diagonal scalar self-energy
    $\widehat{\Pi}_{12,13}^{\lambda_7}$ is inserted into the
    neutral-scalar propagator and the external photon is emitted from
    the top-quark line.
    The right panel shows the type-B contribution involving a
    charged-Higgs loop, with the photon emitted from the charged
    scalar.
    The type-A has a corresponding top-CEDM contribution
    obtained by replacing the photon with a gluon, whereas the type-B does not generate a CEDM because the charged Higgs boson
    is color neutral.
    }
    \label{fig:two_loop_top_edm}
\end{figure}

The representative two-loop top electroweak dipole topologies are
shown in Fig.~\ref{fig:two_loop_top_edm}.
The type-A contribution is generated by inserting the
$\lambda_7$-induced renormalized off-diagonal neutral-scalar
self-energy into the scalar propagator, with the external gauge boson
emitted from the top-quark line.
The type-B contribution instead contains a charged-Higgs loop, with
the external gauge boson emitted from the charged scalar.

After integrating out the additional scalars at $\Lambda$ and the
top quark and electroweak gauge bosons at a lower scale $\mu$, the
electron EDM is evaluated in the leading double-logarithmic
approximation:
\begin{equation}
\begin{aligned}
d_e
={}&
-\frac{e}{2v}
\left(
    \frac{v}{\Lambda}
\right)^2
\left(
    \ln\frac{\Lambda}{\mu}
\right)^2
\nonumber\\
&\times
\left[
    (A_e-D_e)
    \operatorname{Im}C_{tB}
    +
    (B_e-E_e)
    \operatorname{Im}C_{tW}
\right].
\end{aligned}
\label{eq:electron_edm_double_log}
\end{equation}
The coefficients are
\begin{align}
    \mathcal Y_e
    &=
    \frac{
        N_cy_ey_t
    }{
        (4\pi)^4
    },
    \nonumber\\
    A_e
    &=
    \mathcal Y_e
    (15g_1^2+3g_2^2),
    &
    B_e
    &=
    10\mathcal Y_e g_2^2,
    \nonumber\\
    D_e
    &=
    -6\mathcal Y_e g_1^2,
    &
    E_e
    &=
    -5\mathcal Y_e
    (g_1^2+g_2^2).
\label{eq:electron_edm_coefficients}
\end{align}

We generally take $\Lambda$ to be the characteristic heavy-scalar
mass and compare several choices of the lower matching scale,
\begin{align}
    \mu
    =
    v,\quad
    m_t,\quad
    m_Z,
\end{align}
to assess the residual scale dependence. Since
$d_e\propto\ln^2(\Lambda/\mu)$, this uncertainty can be relevant when
the additional scalars lie close to the electroweak scale.

In the top-philic limit,
\begin{align}
    \rho_{ee}=0.
\end{align}
Consequently, the conventional Barr--Zee contribution requiring a
direct nonstandard-Higgs coupling to the electron is absent. The
electron EDM retained here is instead generated through the sequence
\begin{align}
    \text{heavy scalar}
    \longrightarrow
    \text{top electroweak dipole}
    \longrightarrow
    d_e.
\end{align}
\subsection{Matching to Neutron and Proton EDMs}
\label{subsec:hadronic_edms}

The top CEDM generates the Weinberg operator when the top quark is
integrated out. At the top threshold,
\begin{align}
    \frac{\delta w^{(t)}}{g_s}
    =
    \frac{g_s^2}{32\pi^2}
    \frac{\widetilde d_t}{m_t}.
\label{eq:top_threshold_weinberg}
\end{align}

After renormalization-group evolution to the hadronic scale
$\mu=2~\mathrm{GeV}$\cite{edm5,edm6}, the single-source mapping employed in the
numerical analysis\cite{edm7,edm8} is
\begin{align}
    d_u
    &=
    1.8\times10^{-9}\,
    e\,\widetilde d_t,
    &
    d_d
    &=
    -2.0\times10^{-9}\,
    e\,\widetilde d_t,
    \nonumber\\
    \widetilde d_u
    &=
    -8.0\times10^{-9}\,
    \widetilde d_t,
    &
    \widetilde d_d
    &=
    -1.7\times10^{-8}\,
    \widetilde d_t,
    \nonumber\\
    w
    &=
    -1.4\times10^{-5}\,
    \mathrm{GeV}^{-1}
    \widetilde d_t.
\label{eq:top_cedm_low_energy_mapping}
\end{align}
If $\widetilde d_t$ is expressed in $\mathrm{GeV}^{-1}$, the induced
light-quark EDMs and CEDMs in
Eq.~\eqref{eq:top_cedm_low_energy_mapping} have mass dimension
$-1$, while $w$ has mass dimension $-2$.

Using the QCD sum-rule matrix elements\cite{edm9} adopted in this analysis, the
neutron EDM is
\begin{equation}
\begin{aligned}
d_n
={}&
0.73d_d
-
0.18d_u
+
e\left(
    0.20\widetilde d_d
    +
    0.10\widetilde d_u
\right)
\nonumber\\
&+
23\times10^{-3}~\mathrm{GeV}\,
e\,w,
\end{aligned}
\label{eq:neutron_edm_mapping}
\end{equation}
while the proton EDM is
\begin{equation}
\begin{aligned}
d_p
={}&
0.73d_u
-
0.18d_d
-
e\left(
    0.40\widetilde d_u
    +
    0.049\widetilde d_d
\right)
\nonumber\\
&-
33\times10^{-3}~\mathrm{GeV}\,
e\,w.
\end{aligned}
\label{eq:proton_edm_mapping}
\end{equation}
These expressions include the induced light-quark EDMs, light-quark
CEDMs, and the Weinberg operator. In the parameter region considered
here, the Weinberg-operator contribution induced by the top CEDM can
provide an important part of the hadronic EDMs.

The one- and two-loop top CEDMs must be combined coherently at the
amplitude level:
\begin{align}
    \widetilde d_t^{\mathrm{total}}
    =
    \widetilde d_t^{(1)}[J_5]
    +
    \widetilde d_t^{(2)}[J_7].
\label{eq:total_top_cedm}
\end{align}
The low-energy hadronic EDMs are then evaluated from
$\widetilde d_t^{\mathrm{total}}$:
\begin{align}
    d_n^{\mathrm{total}}
    &=
    d_n
    \left[
        \widetilde d_t^{\mathrm{total}}
    \right],
    &
    d_p^{\mathrm{total}}
    &=
    d_p
    \left[
        \widetilde d_t^{\mathrm{total}}
    \right].
\end{align}
Likewise, the electron EDM must be calculated after coherently adding
the electroweak Wilson coefficients,
\begin{align}
    C_{tB}^{\mathrm{total}}
    &=
    C_{tB}^{(1)}
    +
    C_{tB}^{(2)},
    \\
    C_{tW}^{\mathrm{total}}
    &=
    C_{tW}^{(1)}
    +
    C_{tW}^{(2)}.
\end{align}

The $J_5$ and $J_7$ contributions originate from distinct
CP-violating structures:
\begin{align}
    J_5
    &=
    \operatorname{Im}
    (\lambda_5\rho_{tt}^2)
    &&\text{generates the one-loop $H/A$ top dipole},
    \\
    J_7
    &=
    \operatorname{Im}
    (\lambda_7\rho_{tt})
    &&\text{generates the leading two-loop
    scalar-self-energy contribution}.
\end{align}


\end{document}